\documentclass[twocolumn,floatfix
]{revtex4}

\usepackage{amsfonts}
\usepackage{amsmath}
\usepackage{amssymb}
\usepackage{mathrsfs}
\usepackage[english,activeacute]{babel}
\usepackage[latin1]{inputenc}
\usepackage{graphicx}
\usepackage{subfigure}
\usepackage{epsfig}
\usepackage{float}

 \usepackage{color}

\begin{document}

\title{Electromagnetic transmittance in alternating material-metamaterial layered structures}

\author{V.H. Carrera-Escobedo}
\author{H.C. Rosu}\email{hcr@ipicyt.edu.mx}
\affiliation{IPICyT, Instituto Potosino de Investigacion Cientifica y Tecnologica,\\
Camino a la presa San Jos\'e 2055, Col. Lomas 4a Secci\'on, 78216 San Luis Potos\'{\i}, S.L.P., Mexico}%


\begin{abstract}
\noindent Using the transfer matrix method we examine the parametric behavior of the transmittance of $TE$ and $TM$ electromagnetic plane waves propagating in frequency range which are far from the absorption bands of a periodic multilayered system. We focus on the dependence of the transmittance on the frequency and angle of incidence of the electromagnetic wave for the case
in which the periodic structure comprises alternating material-metamaterial layers of various permittivities and permeabilities. A specific example of high transmittance at any angle of incidence in the visible region of the spectrum is identified.\\


%
%
\end{abstract}

\centerline{$\qquad \qquad \qquad \qquad \qquad \qquad \qquad \qquad \qquad \qquad \qquad \qquad \qquad$
Revista Mexicana de F\'{\i}sica 63 (2017) 402-410}

\maketitle

\section{Introduction}

Studies of planar multilayer structures with alternating material and metamaterial layers are
motivated by the known common feature of planar periodic systems to generate transparency bands.
Some time ago, Banerjee {\em et al} \cite{Banerjee::2011}
calculated the intensity of the electromagnetic fields that propagate through 
an array of periodically alternating positive index media (PIM)
and negative index media (NIM). However, they do not
present their results in terms of either the angle of incidence or the frequency of the plane wave, which are the important parameters when one is interested in the directional and frequency selectivities of such structures.
They display the field intensity along the direction of propagation and
compare the transfer matrix method to the finite element method concluding
that the transfer matrix method provides the same results as the latter. Their results motivated us
to use the transfer matrix method
with the main goal of studying the effect of both the angle of incidence and frequency of the propagating plane
wave on the transmittance spectrum.
We compute the values of the transmittance for different values of $\epsilon$ and $\mu$
of the alternating material-metamaterial layers and
search for those values that provide wide windows of transmittance for some regions of frequency
as well as the complementary transmittance gaps that may be useful for making frequency filters \cite{Solymar::2009}.


\section{The transfer matrix method}
\subsection{Waves at an interface}
\begin{figure}[ht] 
\centering 
\includegraphics[width=0.49\linewidth]{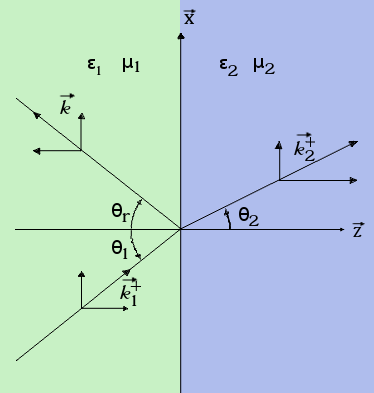}
\caption{Electromagnetic wave vectors at the interface between two different media.}
\label{91}
\end{figure}

To develop the TMM we must first know how the electromagnetic waves behave
at the interface between two dielectrics. For plane waves at an interface,
the electric and magnetic fields are given by
\begin{eqnarray}\label{wi1}
& \vec{E} = \vec{E}_0 e^{i[\vec{k}\cdot\vec{r}-\omega t]}\\
& \vec{H} = \vec{H}_0 e^{i[\vec{k}\cdot\vec{r}-\omega t]}~,
\end{eqnarray}
respectively. The two media separated by the interface will be characterized
by permittivities $\epsilon_1$, $\epsilon_2$ and permeabilities $\mu_1$, $\mu_2$, and the
geometry of the wave vectors at the interface
is illustrated in Fig.~(\ref{91}). In this figure, we assume that the
wave is propagating along the $z$-axis and the wavevector has two components
\begin{equation}\label{wi2}
 \vec{k}=(k_x,0,k_z)~.
\end{equation}

Because the condition of continuity of the tangential components of both the electric and magnetic fields
\begin{equation}\label{wi3}
 \vec{E}_{1t} = \vec{E}_{2t}, ~ \vec{H}_{1t}=\vec{H}_{2t}
\end{equation}
must be satisfied for any point of the interface, the tangential components of $\vec{k}$
have to be equal
\begin{equation}\label{wi4}
 k_{1x}= k_{2x} = k_x~,
\end{equation}
while the dispersion relation $n^2 \omega^2 /c^2 = k^2$ as written for the two media leads to
\begin{eqnarray}\label{wi5}
 \frac{n_1^2 \omega^2}{c^2} = k_1^2= k_x^2 +k_{1z}^2, ~ n_1 = \sqrt{\mu_1 \epsilon_1}~,\\
 \frac{n_2^2 \omega^2}{c^2} = k_2^2= k_x^2 +k_{2z}^2, ~ n_2 = \sqrt{\mu_2 \epsilon_2}~.
\end{eqnarray}
If $k_x$ and $k_z$ are real, 
we can define the angles of incidence and refraction as
\begin{equation}\label{wi6}
\tan \theta_1 = \frac{k_x}{k_{1z}}
\end{equation}
and
\begin{equation}\label{wi7}
 \tan \theta_2 = \frac{k_x}{k_{2z}}~,
\end{equation}
respectively. From Eq.~(\ref{wi4}), we can see that
\begin{equation}\label{wi8}
 k_x = k_1 \sin \theta_1 = k_2 \sin \theta_2
\end{equation}
and if we apply the dispersion relationship, we obtain
\begin{equation}\label{wi9}
 k_x = n_1 \frac{\omega}{c} \sin\theta_1 = n_2 \frac{\omega}{c}\sin\theta_2~.
\end{equation}

If we now take the ratio of the last two equations we obtain Snell's law
\begin{equation}\label{wi}
 \frac{\sin\theta_1}{\sin\theta_2} = \frac{n_2}{n_1}~.
\end{equation}

Considering the case of the TE polarization (s-type polarization), then $\vec{E}$ is parallel
to the interface, {\em i.e.}, we have
\begin{equation}\label{913}
 \vec{E} = (0,E,0)
\end{equation}
and
\begin{equation}\label{914}
 \vec{H}= (H_x,0,H_z)~.
\end{equation}
We also have the boundary condition that says that the component of the electric field which is parallel
to the interface is the same on both sides of the interfaces
\begin{equation}\label{915}
 E_1^+ + E_1^- = E_2^+ + E_2^-~,
\end{equation}
which holds for the magnetic field
\begin{equation}\label{916}
 H_{1x}^+ + H_{1x}^- = H_{2x}^+ +H_{2x}^-
\end{equation}
as well as.

Using Maxwell's equation
\begin{equation}\label{917}
 \vec{k} \times \vec{E} = +\frac{\mu \omega}{c} \vec{H}~,
\end{equation}
in Eq.~(\ref{916}), we obtain
\begin{eqnarray}\label{918}
 \frac{\mu_1 \omega}{c} \vec{H}_{1x}^+ = -k_{1z}E_1^+~,\\
 \frac{\mu_1 \omega}{c} \vec{H}_{1x}^- = k_{1z}E_1^-~.
\end{eqnarray}

Inserting (\ref{918}) in (\ref{916}) leads to
\begin{equation}\label{919}
 -\frac{k_{1z}c}{\mu_1 \omega} E_1^+ + \frac{k_{1z}c}{\mu_1 \omega} E_1^- =
 -\frac{k_{2z}c}{\mu_2 \omega} E_2^+ + \frac{k_{2z}c}{\mu_2 \omega} E_2^-~.
\end{equation}

Now we can write Equations~(\ref{919}) and (\ref{915}) in matrix form
\begin{equation}\label{920}
\left(\begin{array}{cc}
 1 & 1\\
 -\frac{k_{1z}}{\mu_1} & \frac{k_{1z}}{\mu_1}
\end{array}\right)
\left(\begin{array}{c}
 E_1^+\\
 E_1^-
\end{array}\right)
=
\left(\begin{array}{cc}
 1 & 1\\
 -\frac{k_{2z}}{\mu_2} & \frac{k_{2z}}{\mu_2}
\end{array}\right)
\left(\begin{array}{c}
 E_2^+\\
 E_2^-
\end{array}\right)
\end{equation}

or

\begin{equation}\label{921}
\left( \begin{array}{c}
  E_2^+\\
  E_2^-
 \end{array}\right)
= \bf{M^{(s)}}
\left(\begin{array}{c}
 E_1^+\\
 E_1^-
\end{array}\right)
\end{equation}
where the transfer matrix of the interface
\begin{equation}\label{922}
 {\bf{M^{(s)}}}=\frac{1}{2}
\left(\begin{array}{cc}
 1 + \frac{\mu_2 k_{1z}}{\mu_1 k_{2z}} & 1 - \frac{\mu_2 k_{1z}}{\mu_1 k_{2z}}\\
 1 - \frac{\mu_2 k_{1z}}{\mu_1 k_{2z}} & 1 + \frac{\mu_2 k_{1z}}{\mu_1 k_{2z}}
\end{array}\right)~
\end{equation}
is introduced.

By a similar procedure, we can obtain the transfer matrix for the case of the
TM polarization (p-type polarization)
\begin{equation}\label{938}
 \left(\begin{array}{c}
  E_2^+\\
  E_2^-
 \end{array}\right)
= \bf{M^{(p)}}
\left(\begin{array}{c}
 E_1^+\\
 E_1^-
\end{array}\right)
\end{equation}

\begin{equation}\label{939}
 {\bf{M^{(p)}}}=\frac{1}{2}
\left(\begin{array}{cc}
 1 + \frac{\epsilon_2 k_{1z}}{\epsilon_1 k_{2z}} & 1 - \frac{\epsilon_2 k_{1z}}{\epsilon_1 k_{2z}}\\
 1 - \frac{\epsilon_2 k_{1z}}{\epsilon_1 k_{2z}} & 1 + \frac{\epsilon_2 k_{1z}}{\epsilon_1 k_{2z}}
\end{array}\right)~.
\end{equation}

For materials with only real dielectric constants we can express the wave vectors as
\begin{equation}\label{946}
 k_{1z} = k_1 \cos \theta_1~\qquad  {\rm and} ~\qquad  k_{2z}\cos \theta_2~,
\end{equation}
so the transfer matrices for the TE and TM cases are written as
\begin{equation}\label{947}
 {\bf{M^{(s~ or~ p)}}}= \frac{1}{2}
\left(\begin{array}{cc}
 1 + z_{21} \frac{\cos \theta_1}{\cos \theta_2} & 1 - z_{21} \frac{\cos \theta_1}{\cos \theta_2}\\
 1 - z_{21} \frac{\cos \theta_1}{\cos \theta_2} & 1 + z_{21} \frac{\cos \theta_1}{\cos \theta_2}
\end{array}\right)~,
\end{equation}
where $z_{21} = \mu_2 k_1 /\mu_1 k_2$ for the s-polarization and
$z_{21} = \epsilon_2 k_1 /\epsilon_1 k_2$ for the p-polarization.

\subsection{The transfer matrix for a slab}

\begin{figure}[ht]
\centering 
\includegraphics[width=0.49\linewidth]{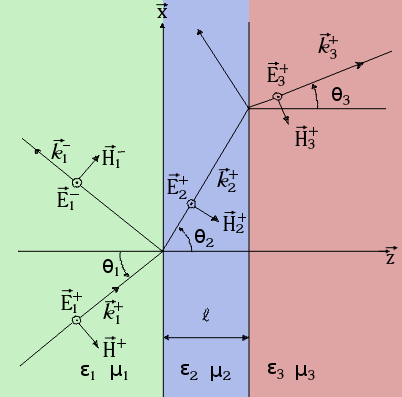}
\caption{Illustration of wave propagation through a slab composed of three different media.}
\label{101}
\end{figure}

For a set of interfaces, the systems behaves as a slab
(sandwich of media). A simple example is given in Fig.~(\ref{101}).
For such systems, it is enough to apply the composition law for transfer matrices \cite{Markos::2008}.
In this way, the relationship of the coefficients of entry and exit is given by
\begin{equation}\label{1001}
\left(\begin{array}{c}
 E_3^+\\
 E_3^-
\end{array}\right)
={\bf M^{12}}
\left(\begin{array}{cc}
 e^{ik_{2z} \ell} & 0\\
 0 & e^{-ik_{2z} \ell}
\end{array}\right)
{\bf M^{23}}
\left(\begin{array}{c}
 E_1^+\\
 E_1^-
\end{array}\right)~,
\end{equation}
where $\ell$ is the length of the slab in the direction of propagation.

From the last equation, one can obtain the transfer matrix for the slab as
\begin{equation}\label{mslab}
 {\bf M_{slab}^{s}} =
 {\bf M^{12}}
\left(\begin{array}{cc}
 e^{ik_{2z} \ell} & 0\\
 0 & e^{-ik_{2z} \ell}
\end{array}\right)
{\bf M^{23}}~,
\end{equation}
which can be used for the generalization to the multilayer case.

\subsection{Wave propagation through a multilayered system}

\begin{figure}[ht]\centering 
\includegraphics[width=0.59\linewidth]{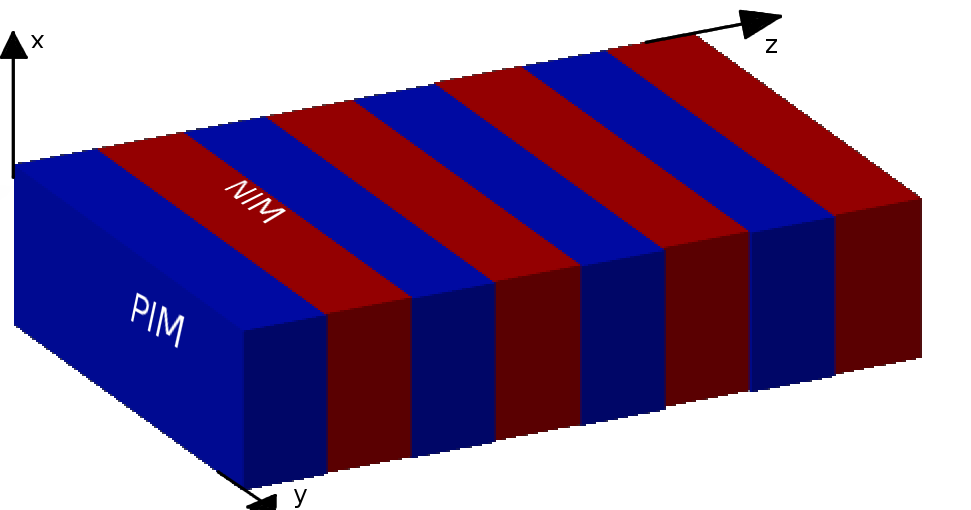}
\caption{A system formed by two types of slabs. The first one is a PIM (blue) labeled as $A$ in the text,
and the second is a NIM (red) labeled as $B$ in the text.}
\label{ML}
\end{figure}

The next step is to consider a multilayered system as illustrated in Fig.~(\ref{ML}).
For a system of this class, the
transfer matrix is obtained by applying the composition law again. In this way, we 
have an \textit{interface} matrix ${\bf M^{i,i+1}}$ for every interface
and a \textit{propagation} matrix of the form
\begin{equation}
{\bf P}=
\left(\begin{array}{cc}
 e^{ik_{2z} \ell} & 0\\
 0 & e^{-ik_{2z} \ell}
\end{array}\right)
\end{equation}
for every slab in which the electromagnetic wave propagates. In this manner, for a ten multilayer slab
system (5 $A$ and 5 $B$ layers) the transfer matrix is written as
\begin{equation}
 {\bf M} = M_{0A} [P_A M_{AB} P_B M_{BA}]^{4} P_A M_{AB} P_B M_{B0}
\end{equation}
where 
 $M_{0X}$ is the interface matrix between air and medium $X$,
 $P_X$ is the propagation matrix of medium $X$, and
 $M_{XY}$ is the interface matrix between medium $X$ and medium $Y$.
The ${\bf M}$ and $M_{XY}$ matrices are transfer matrices for the s-polarization or
the p-polarization, depending on the nature of the incident wave.

Once the transfer matrix for the multilayered system has been defined, one can proceed with the calculation of the transmission
amplitude based on its definition for the case of electromagnetic waves given by \cite{Markos::2008}
\begin{equation}
 t = \frac{det~ {\bf M}}{{\bf M}_{22}}~,
\end{equation}
where $det\,\, {\bf M}$ stands for the determinant of ${\bf M}$, and the transmittance as its square modulus
\begin{equation}
 T = |t|^2~.
\end{equation}

\section{Numerical simulations}

The computation of the transmittance is performed by using a simple python code
\cite{python}.
The system which we model has up to ten slabs, alternating a PIM with a NIM.
The PIM is characterized by $\mu_+$ and $\epsilon_+$, while the NIM by $\mu_-$ and $\epsilon_-$.
The frequency of the incident wave is in $c/(\ell \sqrt{\epsilon_0})$ units, where
the chosen numerical value for $\ell$ is 1.0 $\mu$m, and the angle of
incidence $\theta$ goes from $-\pi /2 < \theta < \pi /2$ which we normalize to
$\theta_0=\pi /2$ in the figures. The plots of the transmittance
versus frequency are obtained for an angle of incidence of $\pi/3$ radians.
The range of frequencies goes from 300 THz to 1200 THz, which entails the frequency
band of the visible light (430-770 THz). In the units of the plots, the band of frequencies
of the visible light is [1.4-2.5] and is indicated by two horizontal black solid lines
in the contour figures.

\subsection{The effect of the number of layers}

\begin{figure}[ht]\centering
\includegraphics[width=0.49\linewidth]{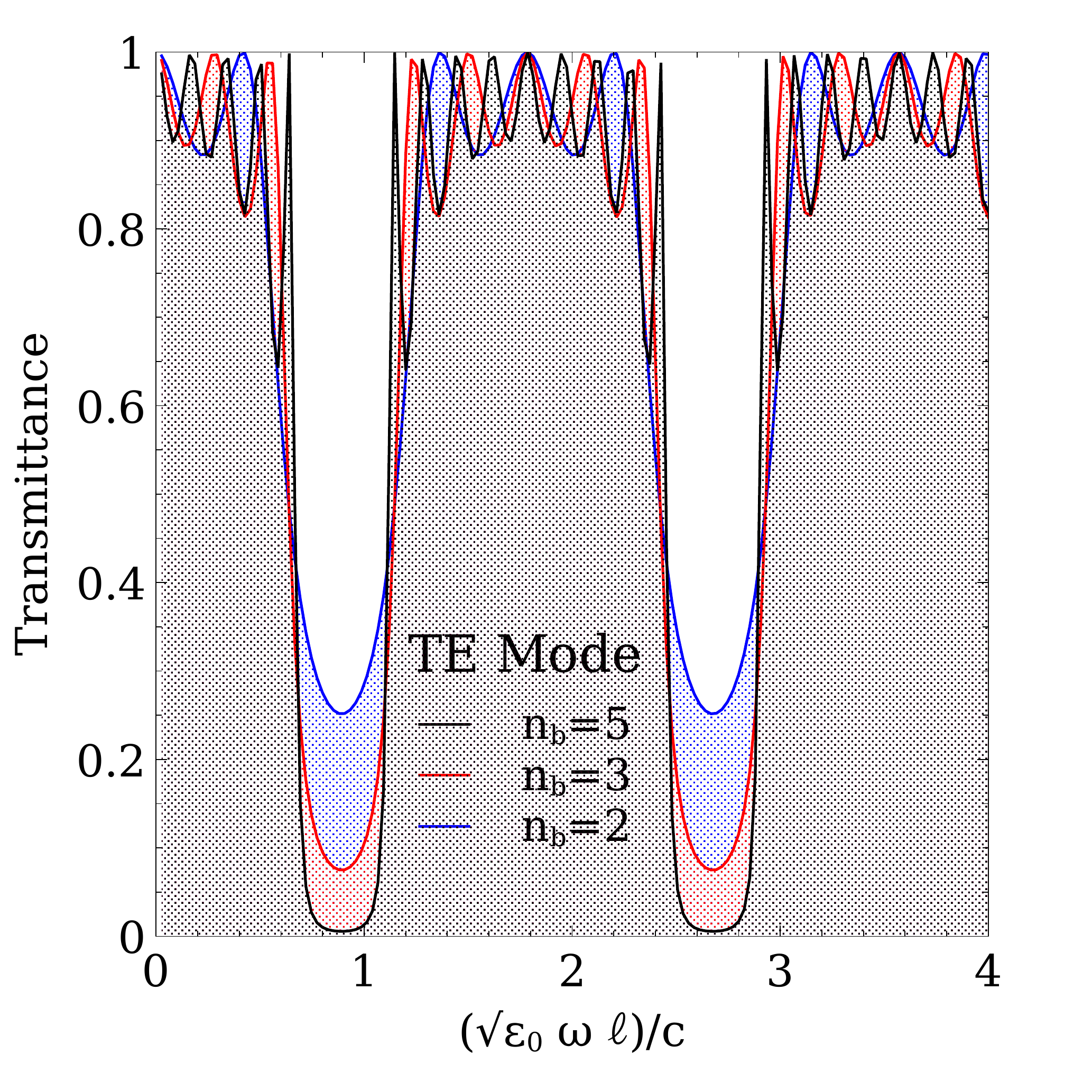}
\includegraphics[width=0.49\linewidth]{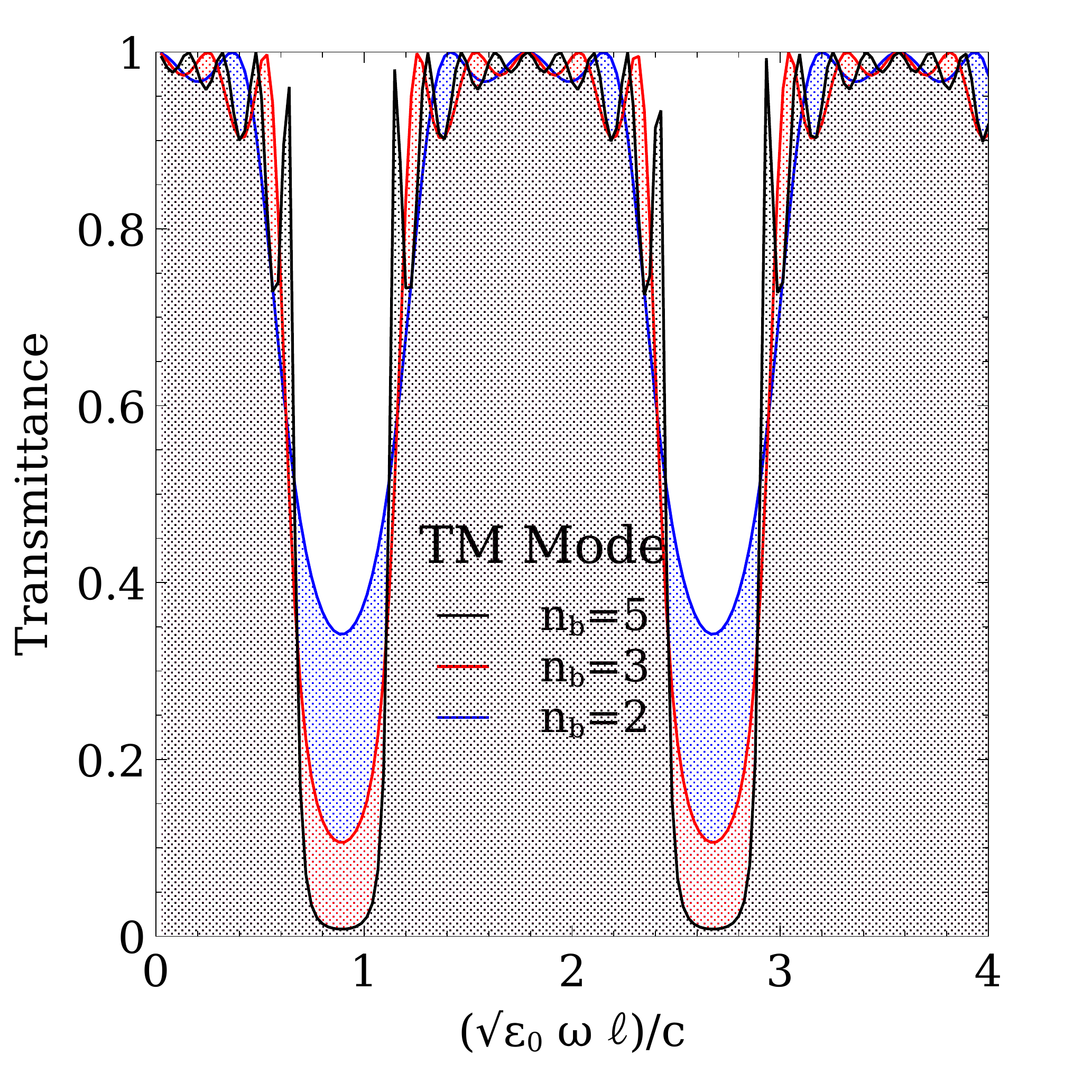}
\caption{Plots corresponding to the case of variable number of blocks.
Left panel: the TE mode. Right panel: the TM mode.
In these graphs we can see that the valleys of the transmittance get deeper
as we increase the number of blocks. This is congruent with the formation
of transmission bands in the case of superlattices.}
\label{nb-sl}
\end{figure}

\begin{figure*}[ht]\centering 
\includegraphics[width=0.30\linewidth]{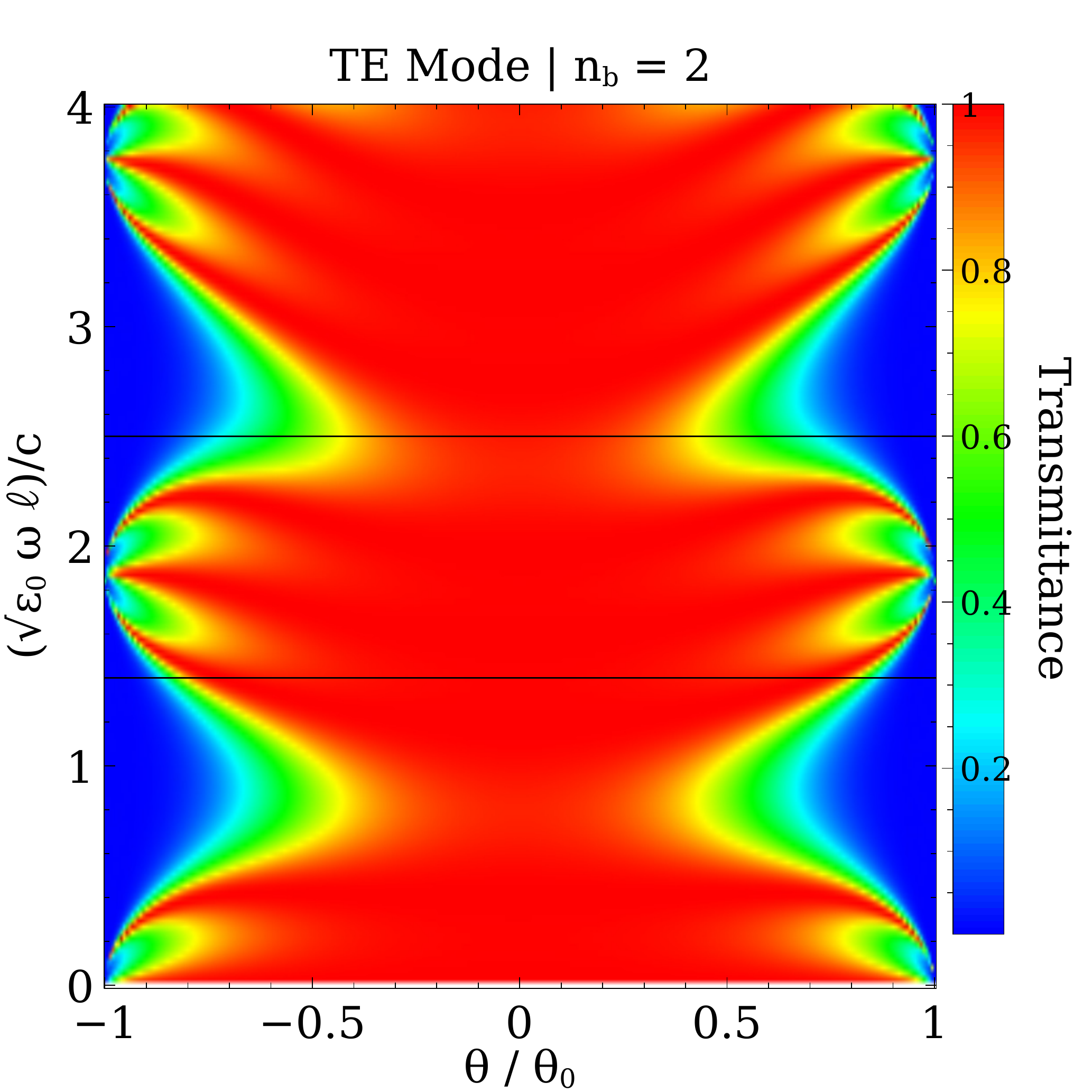}
\includegraphics[width=0.30\linewidth]{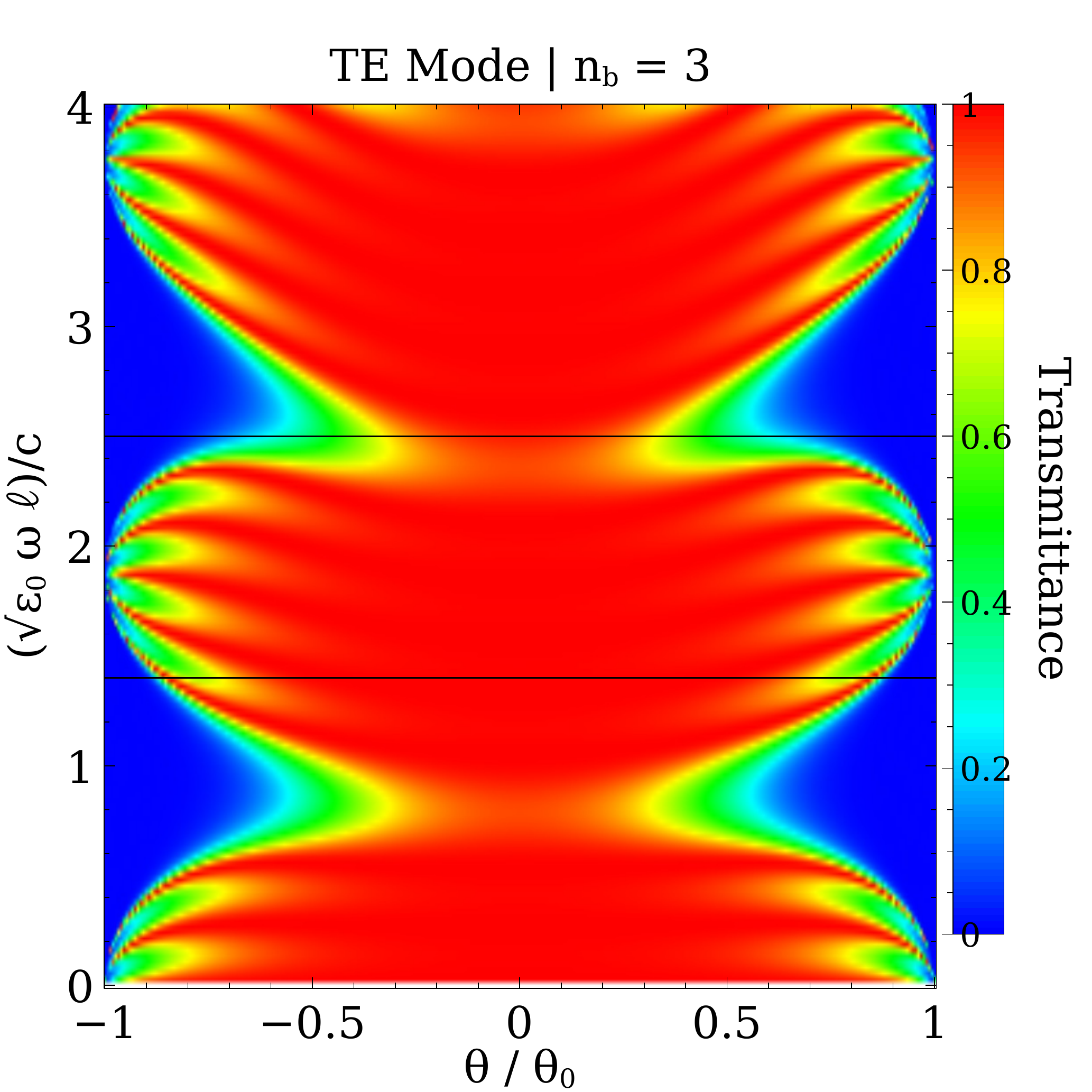}
\includegraphics[width=0.30\linewidth]{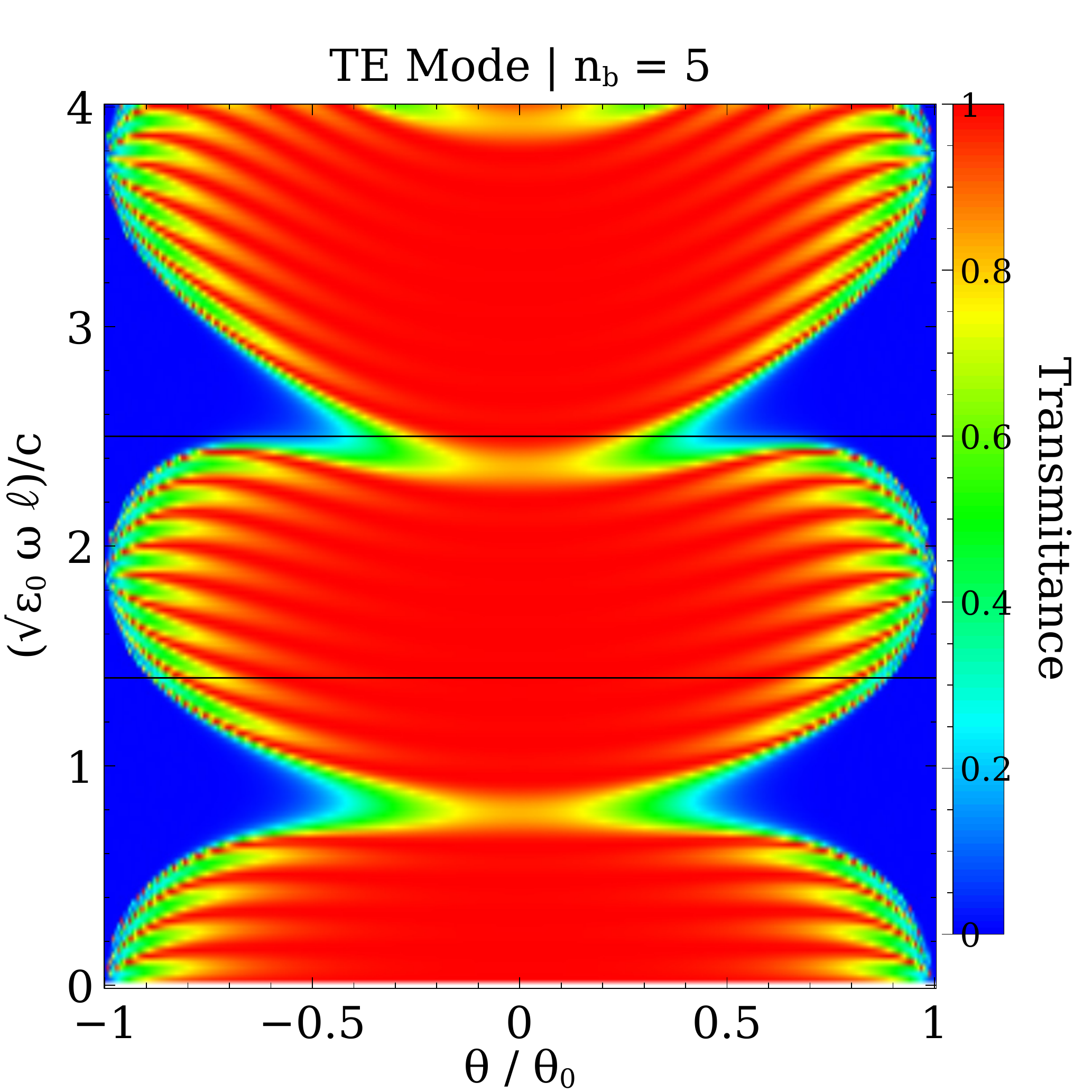}
\includegraphics[width=0.30\linewidth]{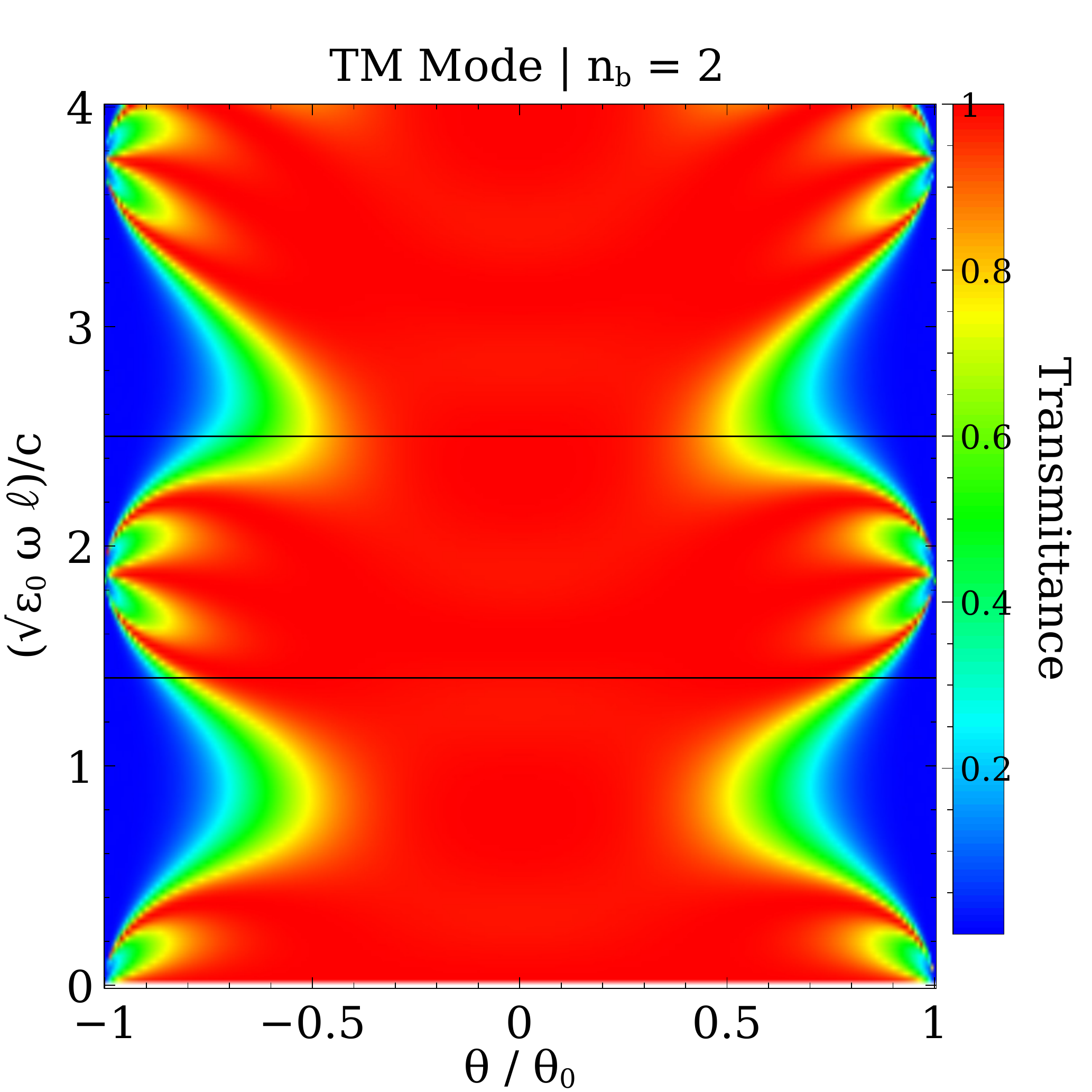}
\includegraphics[width=0.30\linewidth]{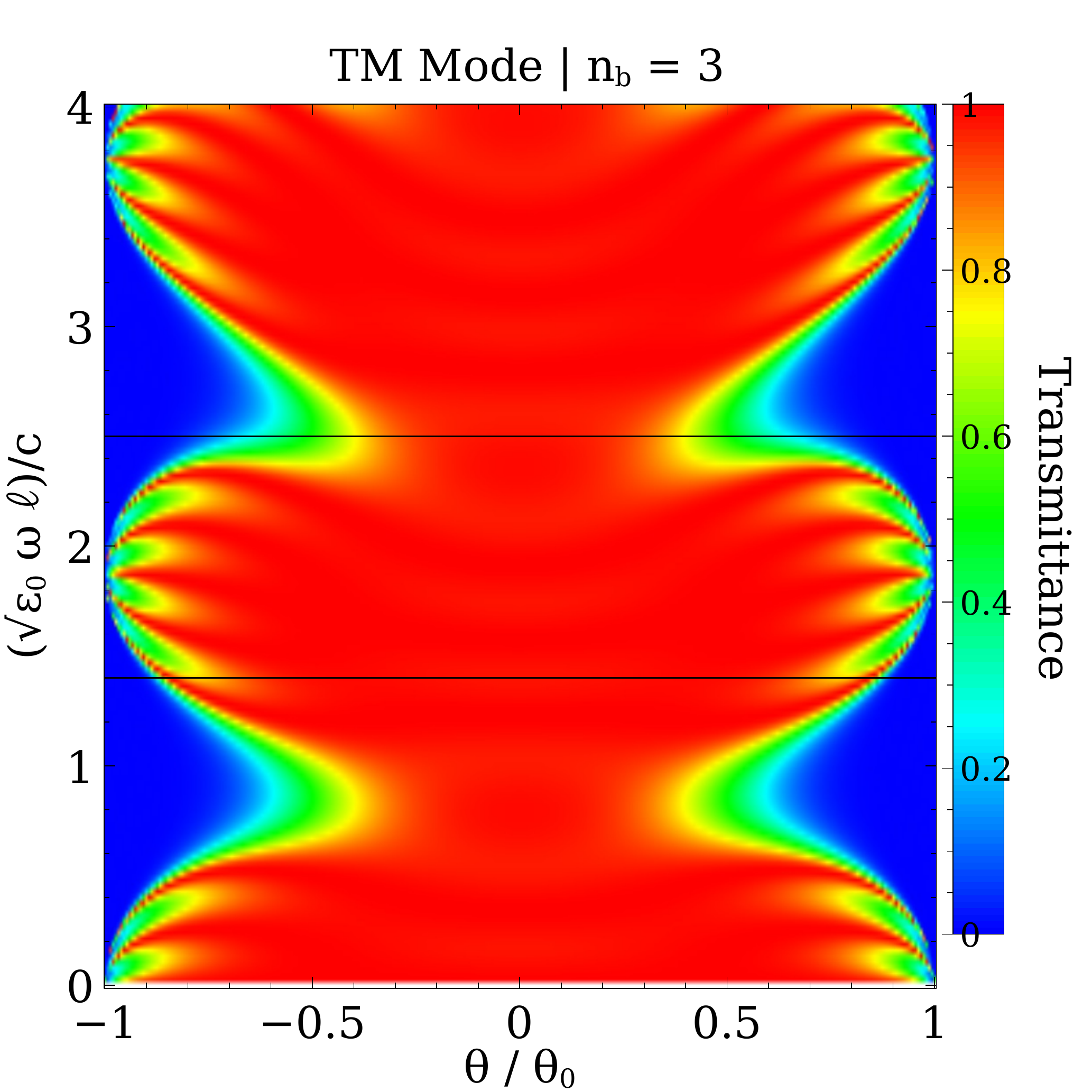}
\includegraphics[width=0.30\linewidth]{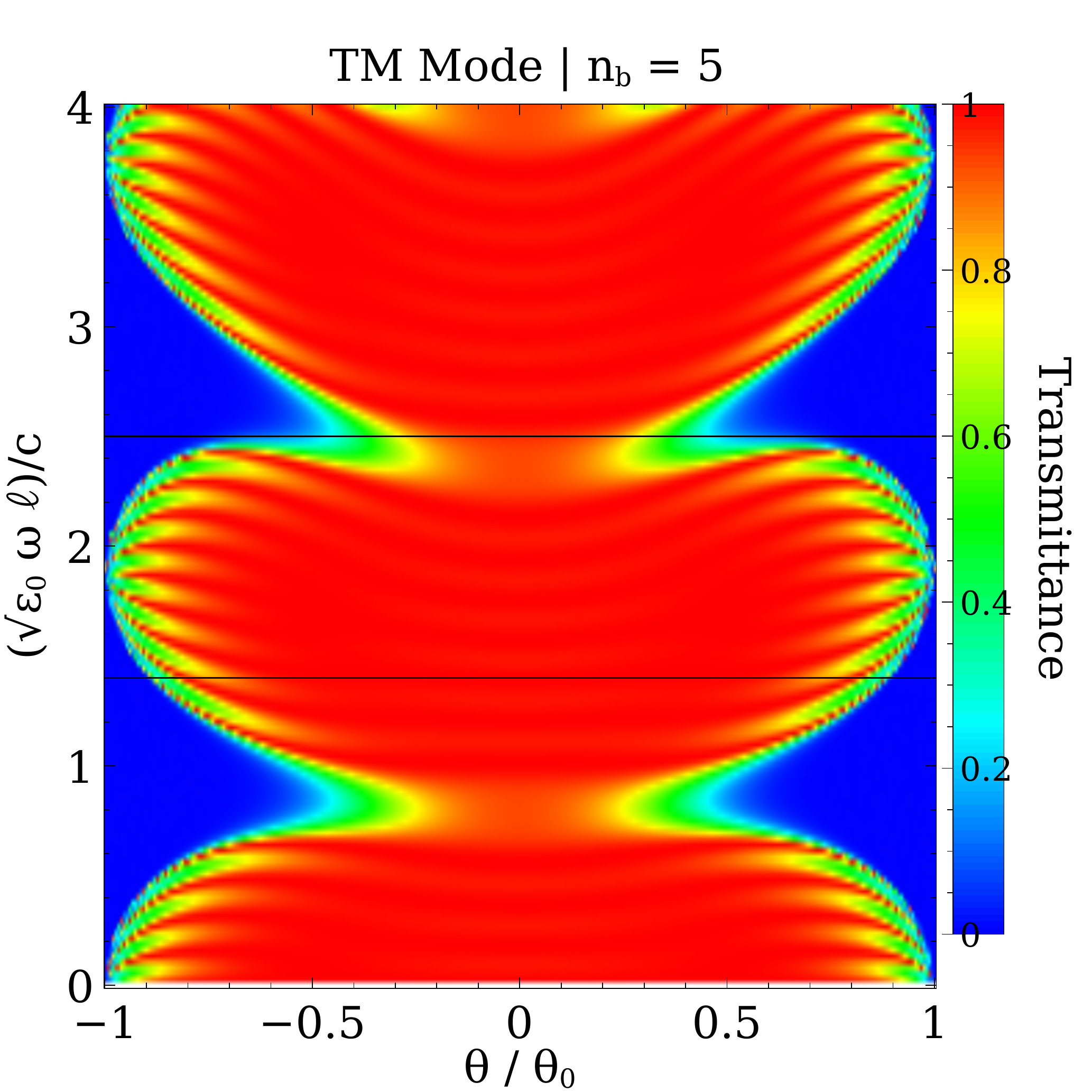}
\caption{Contour plots. The top row corresponds to the TE mode (s-polarization) while the
bottom row correspond to the TM mode (p-polarization). The number of blocks takes the values
$n_b = 2,3,5$. The $x$-axis corresponds to the angle of incidence, the $y$-axis to the
frequency of incidence in $\sqrt{\epsilon_0} \ell / c$ units, and the color bar corresponds to
transmittance. The black horizontal lines indicate the limits of the frequencies of the visible
range.}
\label{ctnb}
\end{figure*}

We begin by analyzing the effect of the number of layers upon the
transmission spectrum. With this task in mind, we vary the number of periods ``AB''
crossed by the propagating wave using the values of the parameters
given in Table~\ref{tabla1},
\begin{table}
\centering
\begin{tabular}{|l|l|}\hline
Parameter & Value\\\hline
$\epsilon_0$ & 1.0\\\hline
$\epsilon_+$ & 2.0\\\hline
$\epsilon_-$ & -1.0\\\hline
$\mu_0$ & 1.0\\\hline
$\mu_+$ & 2.0\\\hline
$\mu_-$ & -1.2\\\hline
$n_b$ & 2,3,5\\\hline
\end{tabular}
\caption{The values of the permeabilities and permittivities of the AB slabs used for the case
of a variable number of blocks.}
\label{tabla1}
\end{table}
where the parameters $\epsilon_0$ and $\mu_0$ correspond to the respective values for the
relative permittivity and permittivity of the medium (usually air) in front and at the end of
the ``AB'' multilayer structure. We call the ``AB'' period a block. In Fig.~(\ref{ctnb}), the
number of blocks is increased from two to five, (in terms of interfaces, from five to eleven),
and each panel is labeled by the corresponding number of blocks.
The regions of high transmittance show up as three symmetrical
bubbles. Upon increasing the number of blocks in the system, one can see that the number of peaks inside
each of the bubbles increases. The number of peaks, $n_{peaks}$, is equal to twice the number of blocks, $n_{b}$,
minus one, $n_{peaks}=2n_{b}-1$. This might be seen alike to the case of quantum systems if the number of quantum wells
is put in correspondence with the number of peaks in the transmittance.
One can also see that when the number of blocks increases the transmittance between the bubbles
of high transmittance becomes smaller. This is a consequence of the change in the transmittance
values that one can see in Fig.~(\ref{nb-sl}), where we can also notice that the transmittance bands
get wider as the number of blocks increases.
Moreover, in Fig.~(\ref{ctnb}), we can see that we have a bubble of high transmittance inside
the limits of the visible frequencies (black horizontal lines), which could be useful for camouflage purposes.
In addition, as we increase the number of blocks, this high transmittance region entails a wider range of angular values.
On the other hand, if we increase the number of blocks enough the simulation ends up in a completely opaque
material with zero transmittance, as expected.

\subsection{The effect of $\epsilon_+$ and $\epsilon_-$}

\begin{figure}[ht]\centering
\includegraphics[width=0.49\linewidth]{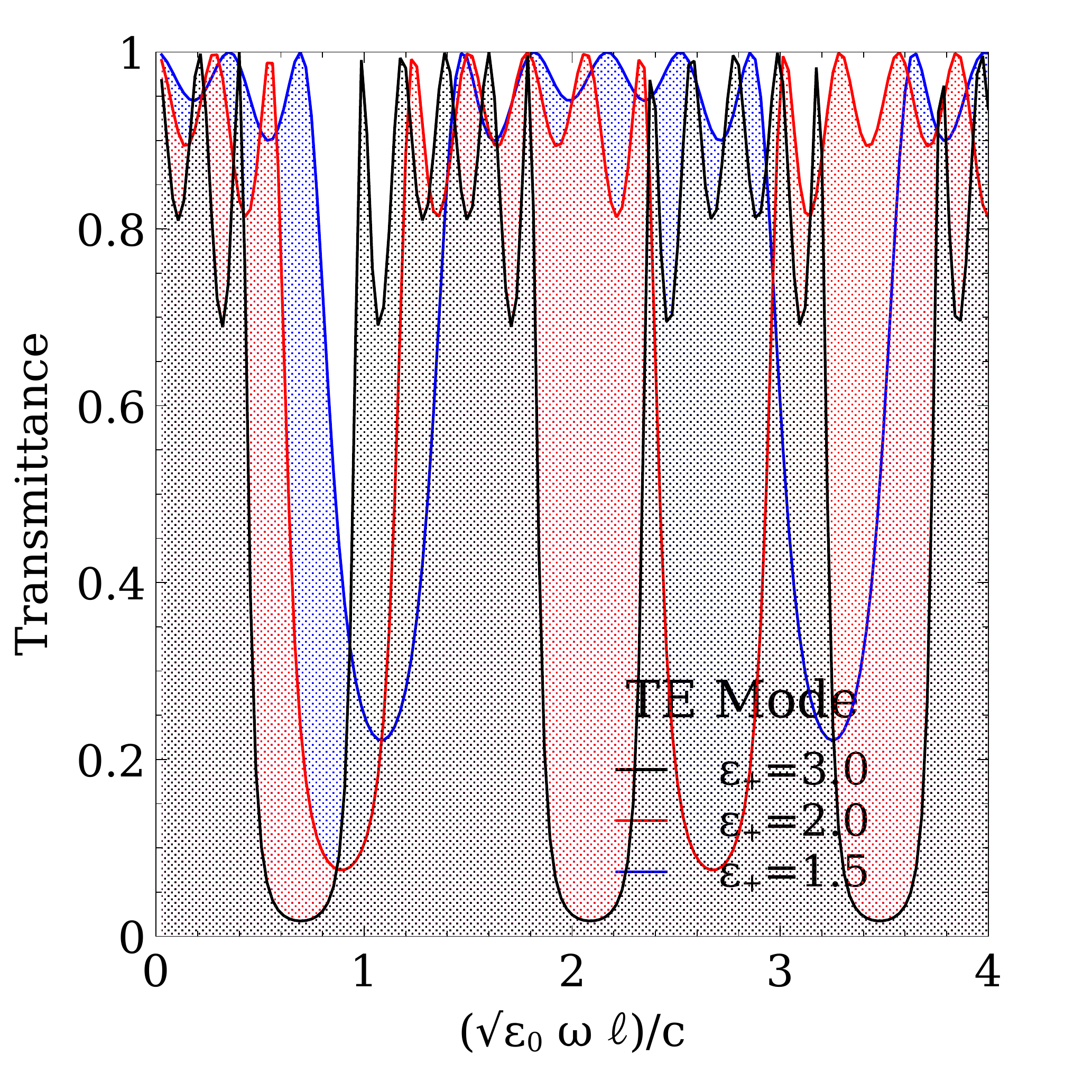}
\includegraphics[width=0.49\linewidth]{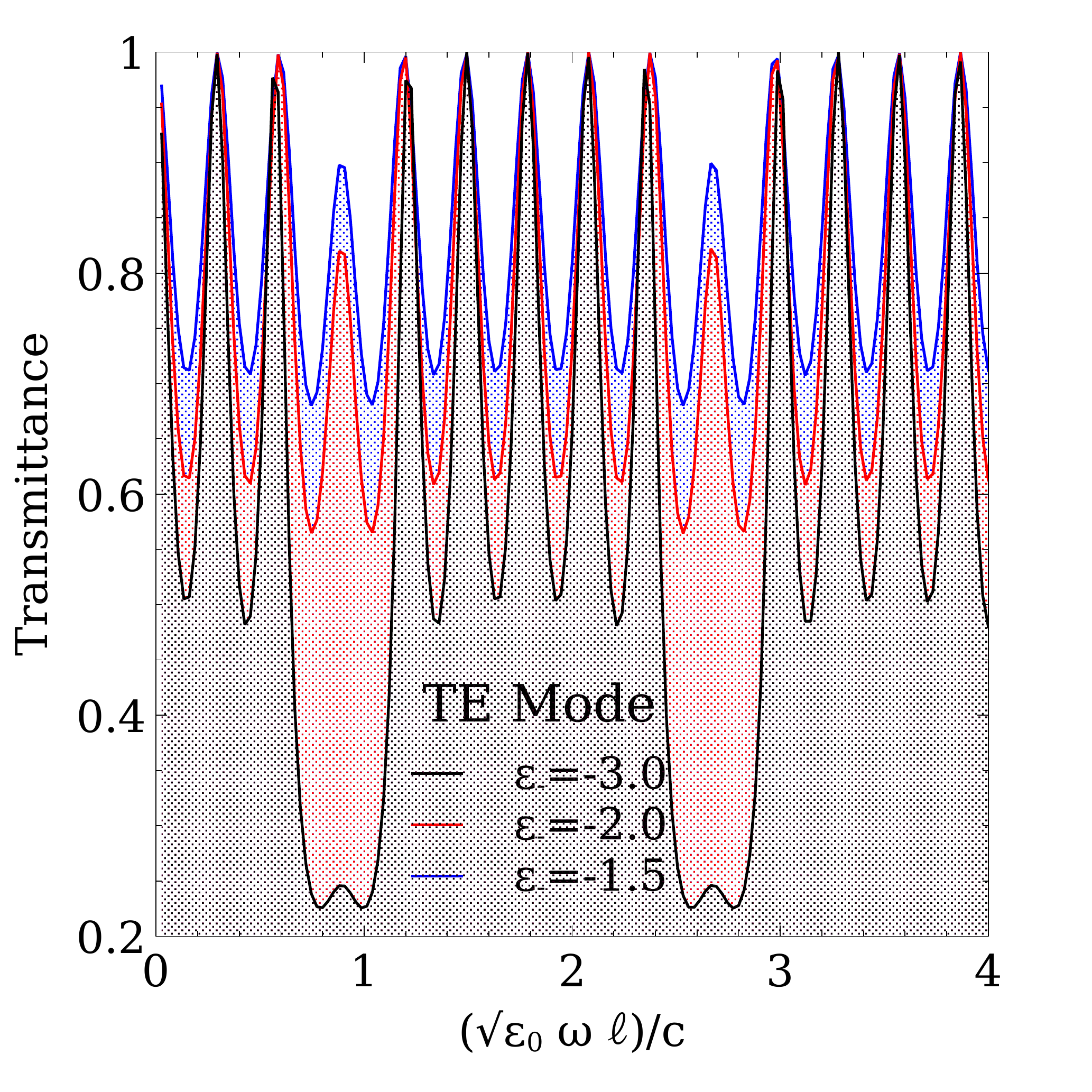}
\caption{Plots corresponding to the case of fixed angle of incidence
$\theta=\pi/3$. The left panel corresponds to the case of variable $\epsilon_+$ and
the right panel corresponds to the case of variable $\epsilon_-$.
In these graphs, we can see the shifting of the peaks in
transmittance only for the case with increasing $\epsilon_+$.}
\label{shift-e}
\end{figure}

Knowing that the number of blocks generates a series of transmittance bands,
one naturally may ask if there is a way to control the width
of these bands by means of the electric permittivities of the slabs.
The values of the parameters that we use for this computation
are given in Table~\ref{tabla2}.
\begin{table}
\centering
\begin{tabular}{|l|l|}\hline
Parameter & Value\\\hline
$\epsilon_0$ & 1.0\\\hline
$\epsilon_+$ & 1.5, 2.0, 3.0\\\hline
$\epsilon_-$ & -1.5, -2.0, -3.0\\\hline
$\mu_0$ & 1.0\\\hline
$\mu_+$ & 2.0\\\hline
$\mu_-$ & -1.2\\\hline
$n_b$ & 3\\\hline
\end{tabular}
\caption{The values of the permeability and permittivity used when the number of blocks is kept fixed.}
\label{tabla2}
\end{table}

The results are presented in Fig.~(\ref{epv}). We can see that the effect of increasing
$\epsilon_+$ while keeping $\epsilon_-$ fixed at $-1.0$,
is to tighten the bubbles of high transmittance, and the magnitude of this effect
is proportional to the increase in $\epsilon_+$. When $\epsilon_+$ is high enough, the bubbles
lose the connection between them and become independent ovals. For this case, we see that
the effect is similar for both the TE and TM modes, but for the TM mode the
bubbles are more separated and there appears a series of smaller bubbles of high transmittance.
Moreover, for both types of modes there is a shift to lower frequencies
of these bubbles of high transmittance. This can be seen in the left panel of Fig.~(\ref{shift-e}).

On the other hand, if we increase the value of $|\epsilon_-|$
while keeping $\epsilon_+$ fixed at 2.0 we can see from Fig.~(\ref{emv}) that
the bubbles of high transmittance are absent and in their place a wide band with
horizontal spikes occurs. Therefore, we conclude that the occurrence of bubbles depend on the chosen values
of $\epsilon_+$. When $\epsilon_-$ increases, we can see that the wide band is fragmented
in three sections, and the separation between sections is proportional to the increment in
$|\epsilon_-|$. In this case, there is no shift to lower frequencies. This fact can be observed
in the right panel of Fig.~(\ref{shift-e}).
The effect is similar for both TE and TM modes, but in the TM case the
decrease in transmittance is less but in a wider range of frequencies.

In general, we can say that the value of $\epsilon_+$ determines the form of the transmittance bands and
the frequency range where they occur. On the other hand, the value of $\epsilon_-$ determines the
width of the horizontal transmittance spikes.

\subsection{The effect of $\mu_+$ and $\mu_-$}

\begin{figure}[ht]\centering 
\includegraphics[width=0.49\linewidth]{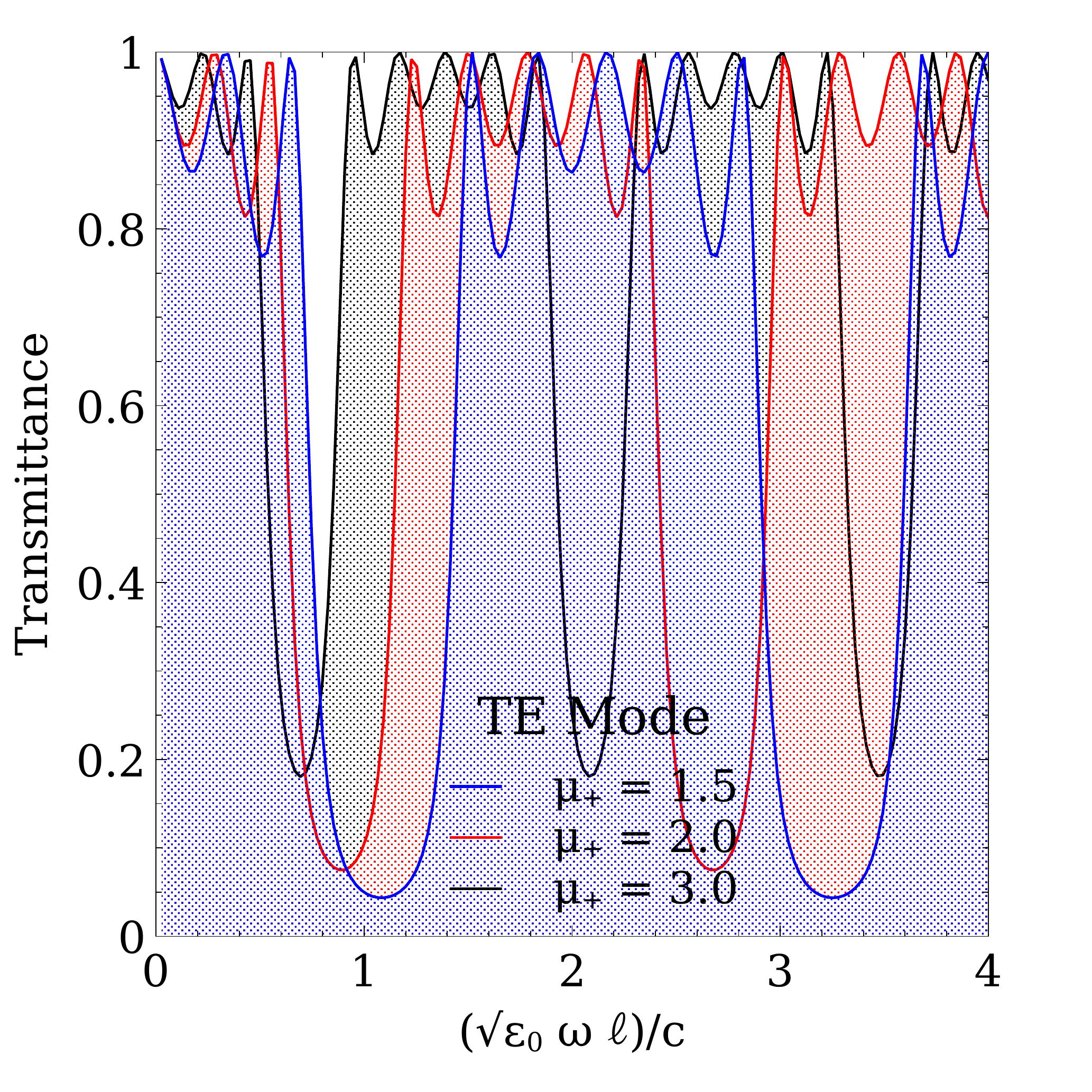}
\includegraphics[width=0.49\linewidth]{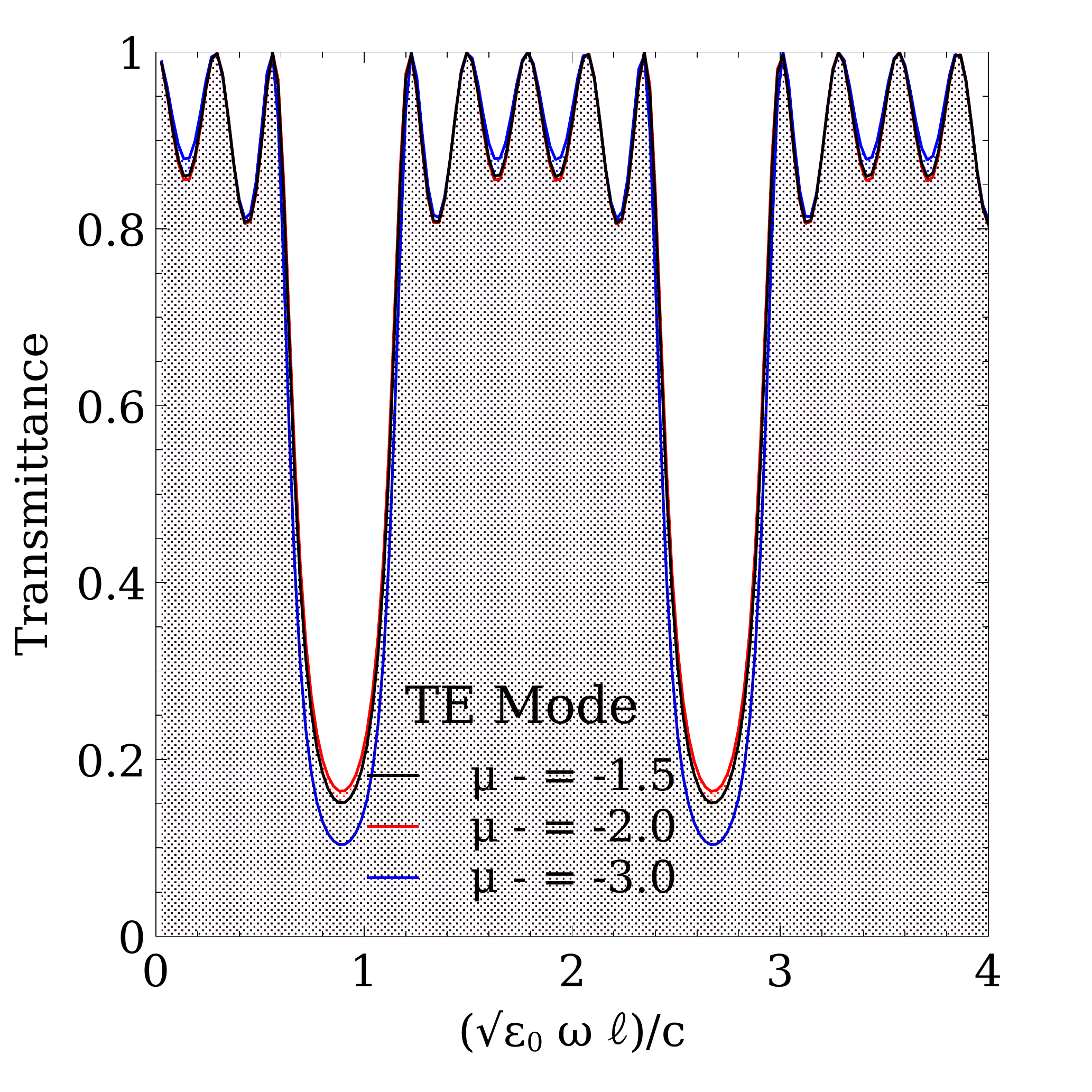}
\caption{Plots corresponding to the case of fixed angle of incidence
$\theta=\pi/3$. Left panel: case of variable $\mu_+$. Right panel:
case of variable $\mu_-$. One can see that the shifting of the peaks in
transmittance occurs only upon increasing $\mu_+$ (left panel) and
there is no shifting effect when $|\mu_-|$ is increased (right panel).}
\label{shift-m}
\end{figure}

We move now to the study of the effects of increasing the values of the permeabilities of the system.
For this task, we use the values from Table~\ref{tabla3} for the parameters under control.

\begin{table}
\centering
\begin{tabular}{|l|l|}\hline
Parameter & Value\\\hline
$\epsilon_0$ & 1.0\\\hline
$\epsilon_+$ & 2.0\\\hline
$\epsilon_-$ & -1.0\\\hline
$\mu_0$ & 1.0\\\hline
$\mu_+$ & 1.5, 2.0, 3.0\\\hline
$\mu_-$ & -1.5, 2.0, 3.0\\\hline
$n_b$ & 5\\\hline
\end{tabular}
\caption{The parameters used when the $\mu$ values are changed.}
\label{tabla3}
\end{table}

\begin{figure*}[ht]\centering 
\includegraphics[width=0.30\linewidth]{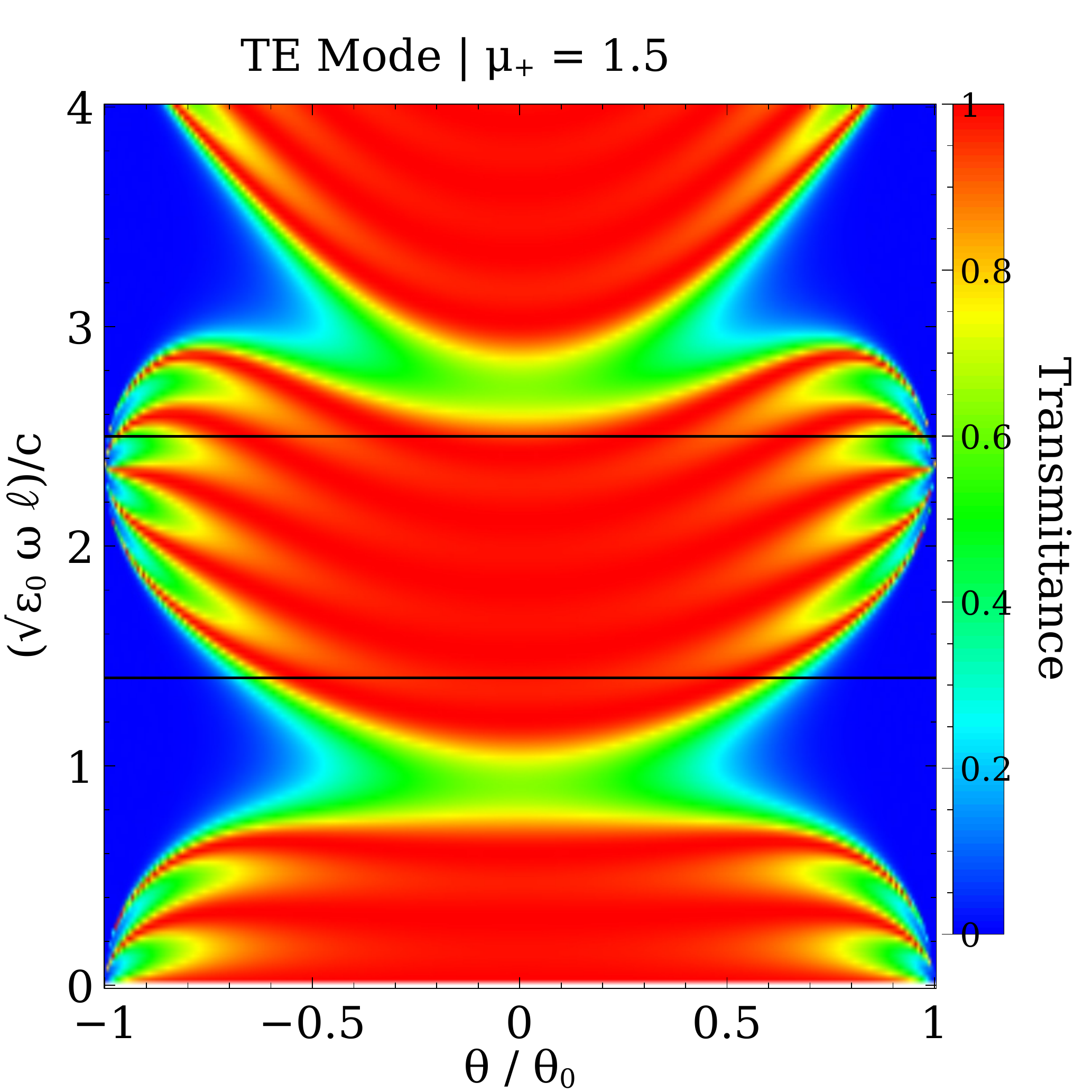}
\includegraphics[width=0.30\linewidth]{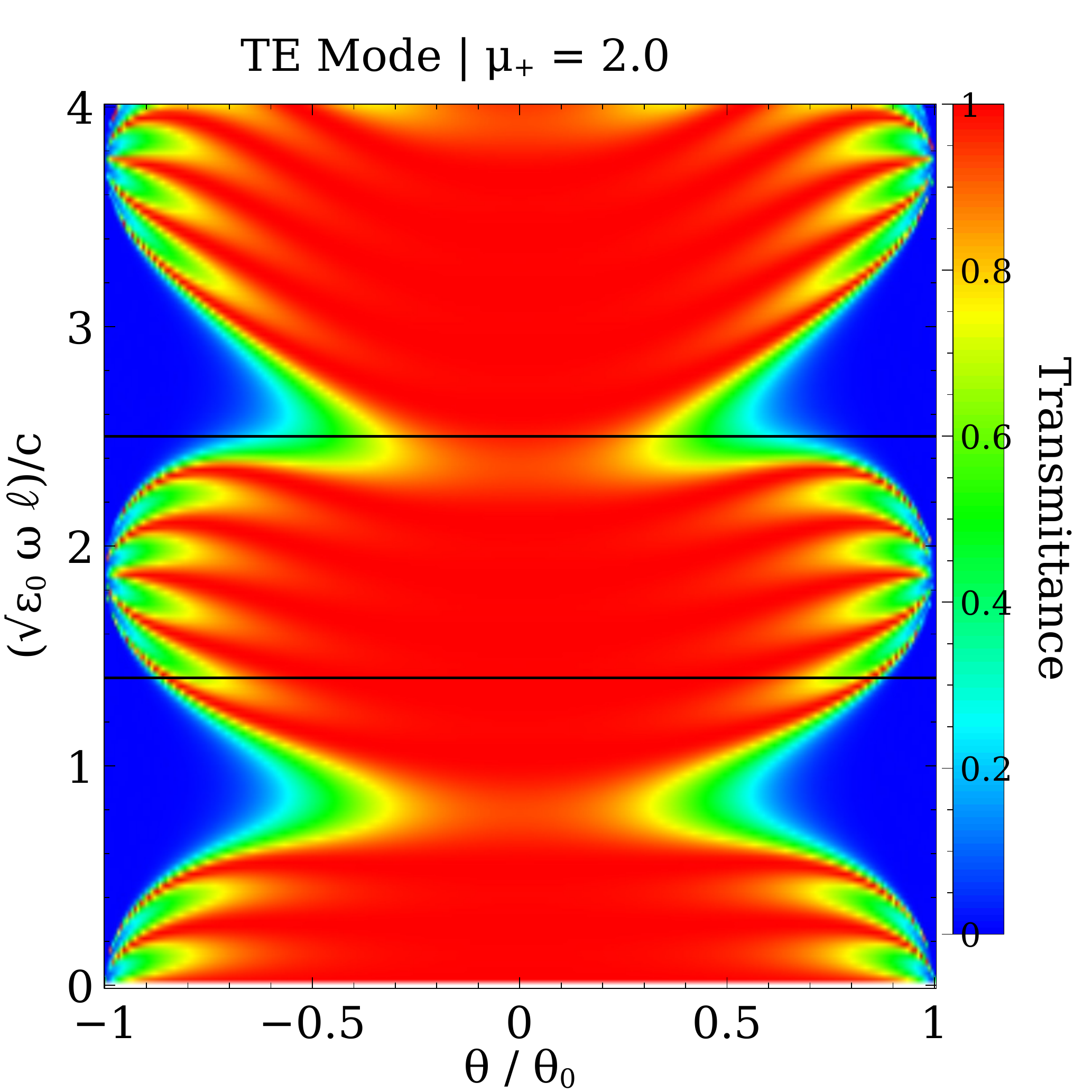}
\includegraphics[width=0.30\linewidth]{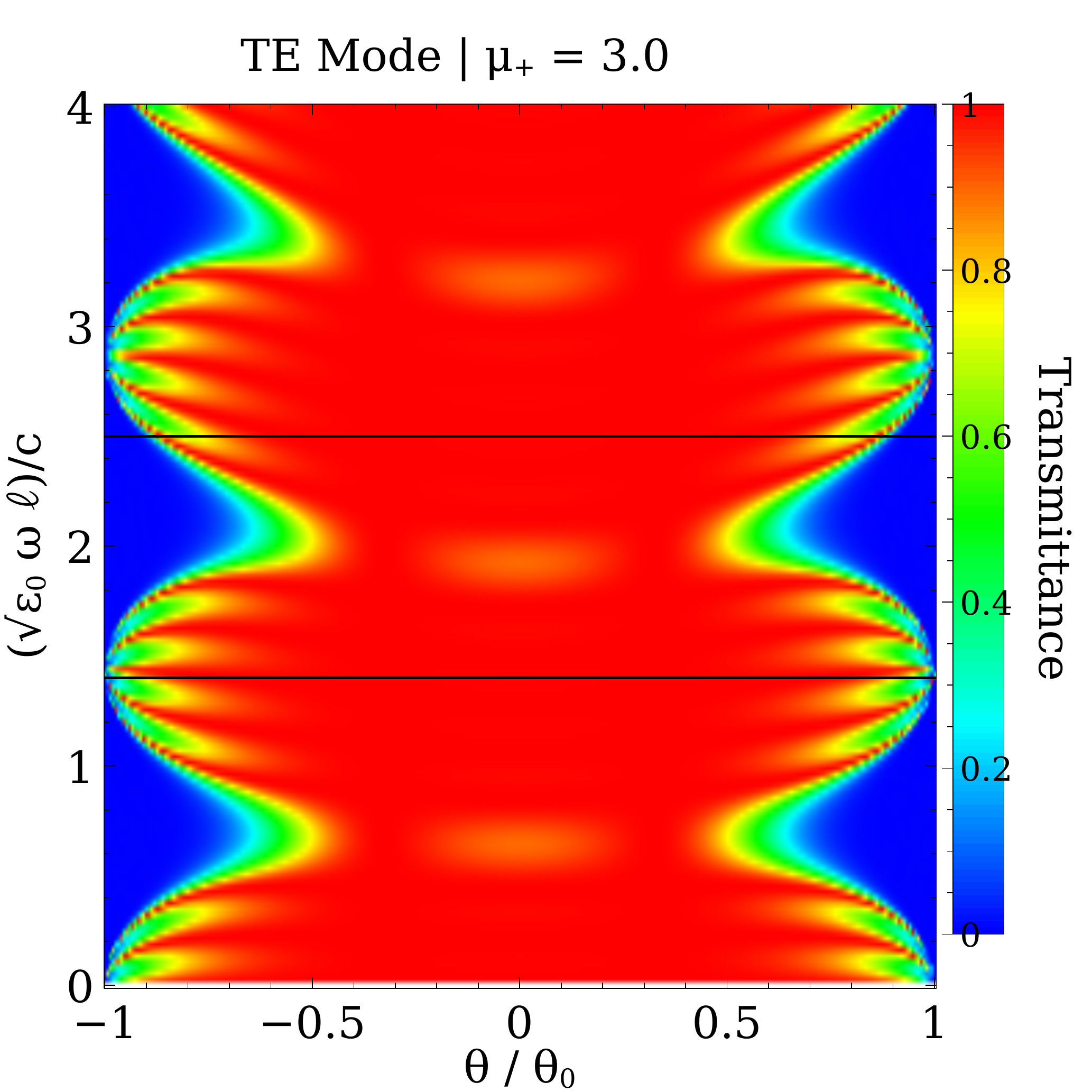}
\includegraphics[width=0.30\linewidth]{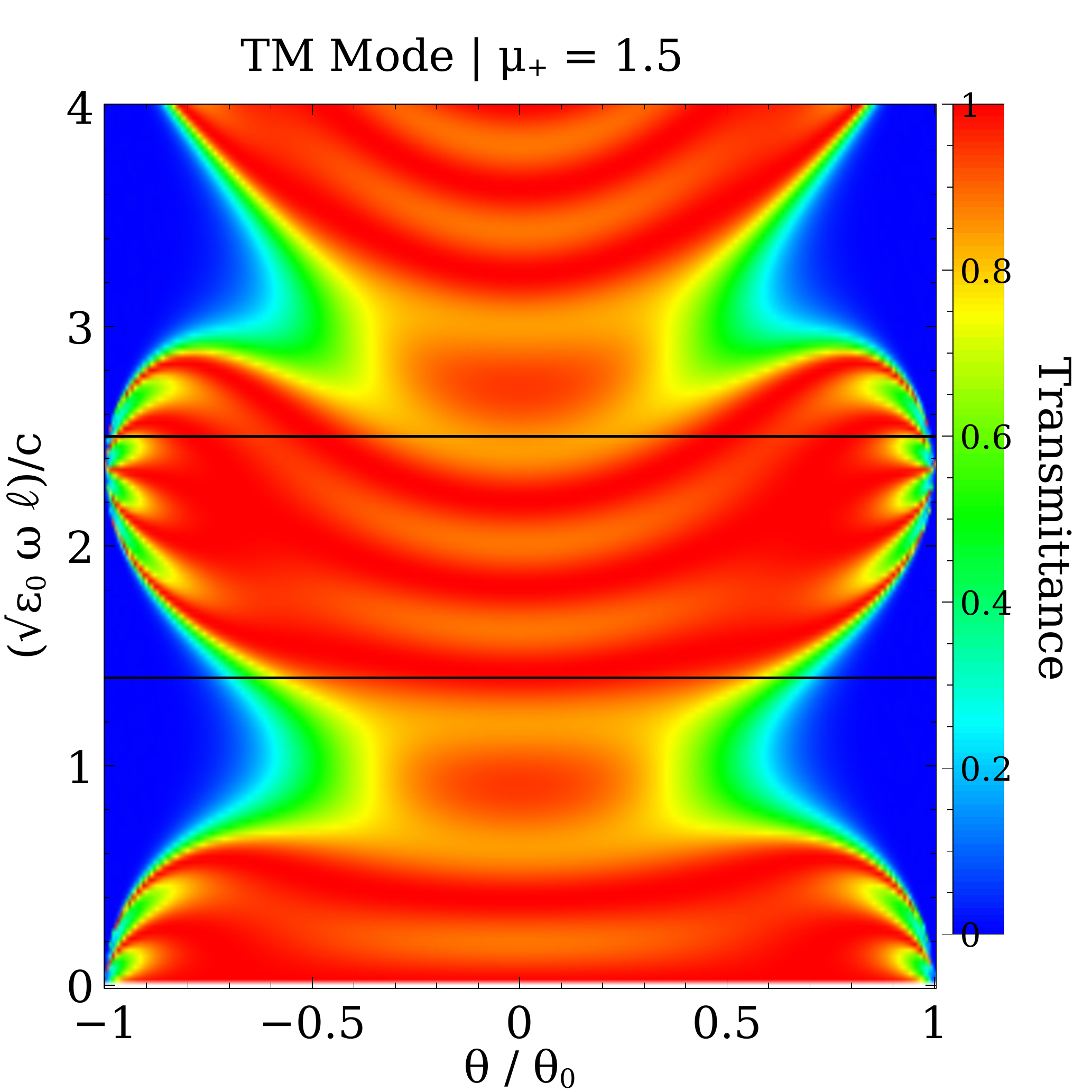}
\includegraphics[width=0.30\linewidth]{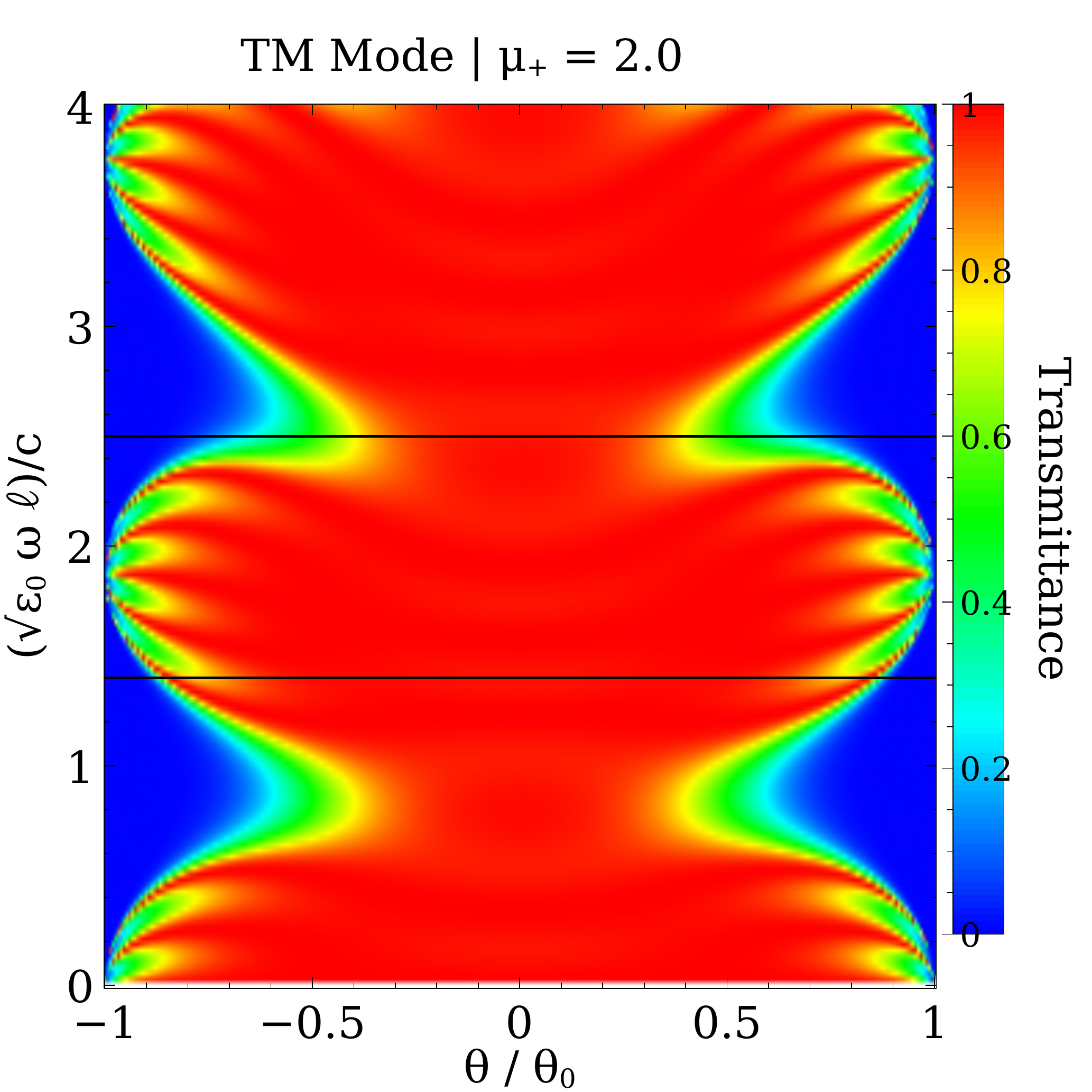}
\includegraphics[width=0.30\linewidth]{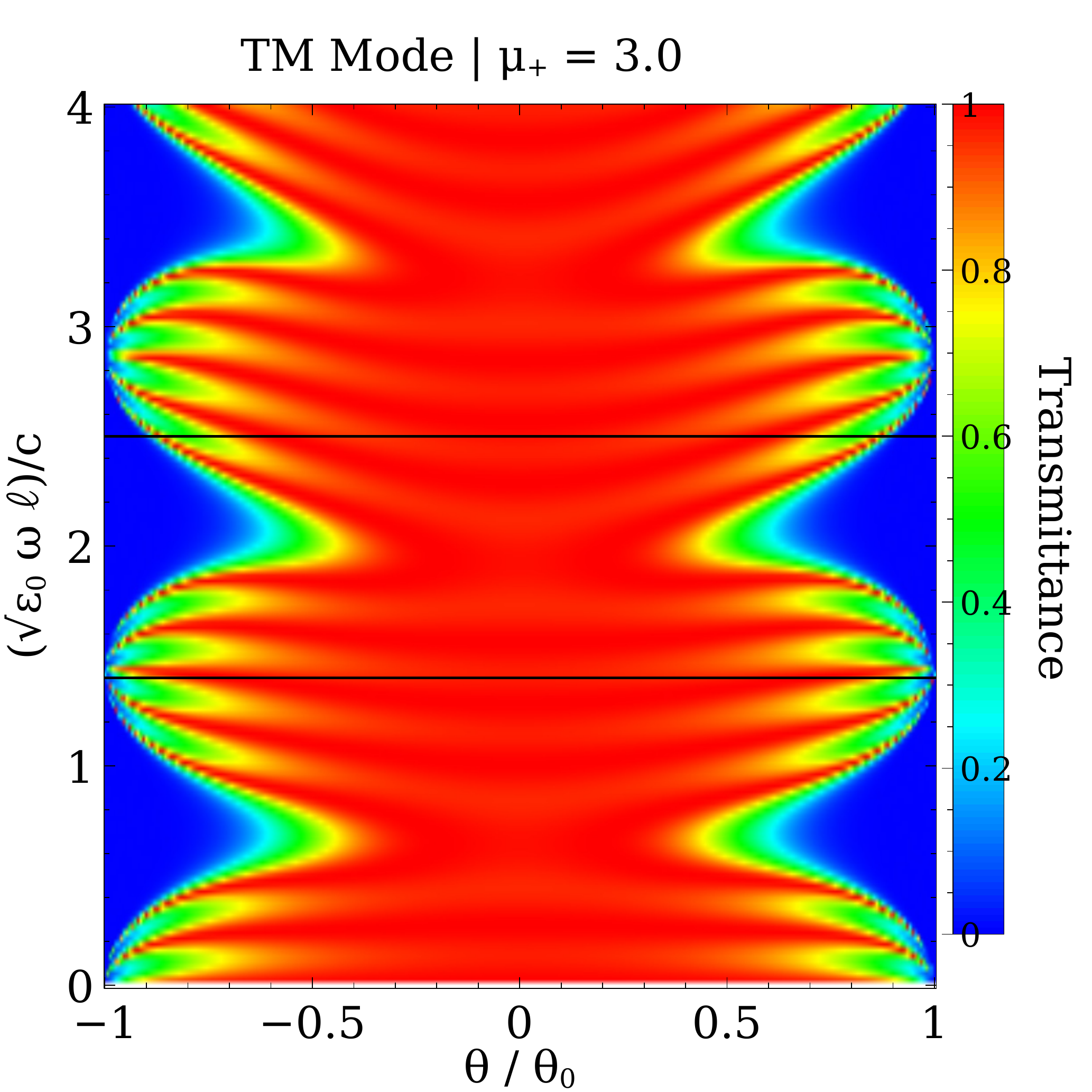}
\caption{Contour plots. The top row corresponds to the TE mode while the
bottom row corresponds to the TM mode. $\mu_-$ is fixed at -1.2.}
\label{mpv}
\end{figure*}

From Fig.~(\ref{mpv}), we see that if we increase the magnitude of $\mu_+$
the bubbles of high transmittance have a shift to lower frequencies which
is proportional to the magnitude of $\mu_+$. Unlike the case of varying
$\epsilon_+$, we see that the bubbles remain connected, and as the value
of $\mu_+$ increases the contact area between bubbles increases as well as.
This is opposite to the case of varying $\epsilon_+$ and occurs for both
the TE and TM modes.

From the left panel of Fig.~(\ref{shift-m}), we can appreciate easily a shifting
effect to lower frequencies. Also, and even more interesting, we see that if we increase the value
of $\mu_+$ the valleys (regions of low transmittance) become less apparent.
Thus, increasing $\mu_+$ has the effect of rising the valleys of transmittance diminishing the relative borders
of the transmittance bands.

\begin{figure*}[ht]\centering 
\includegraphics[width=0.30\linewidth]{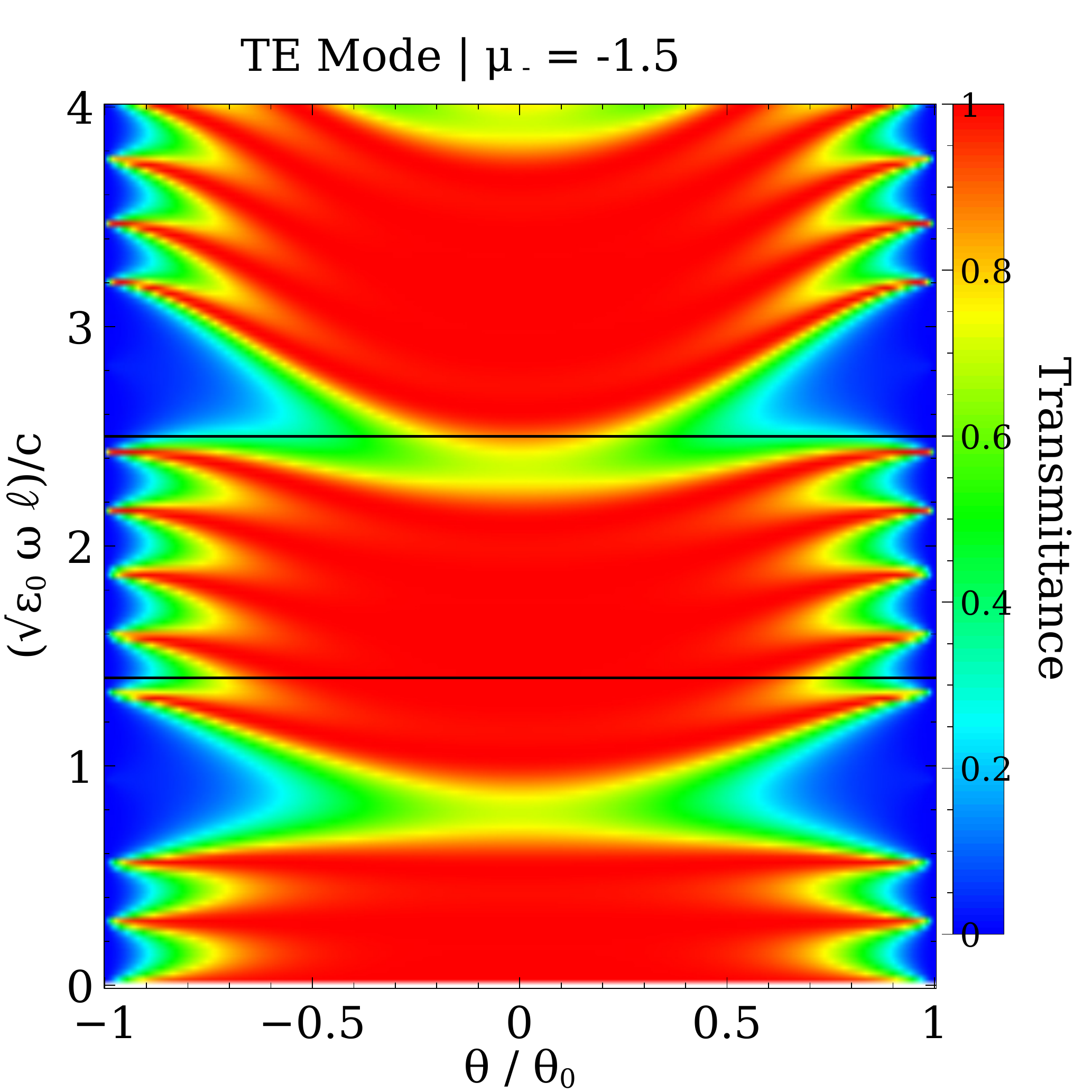}
\includegraphics[width=0.30\linewidth]{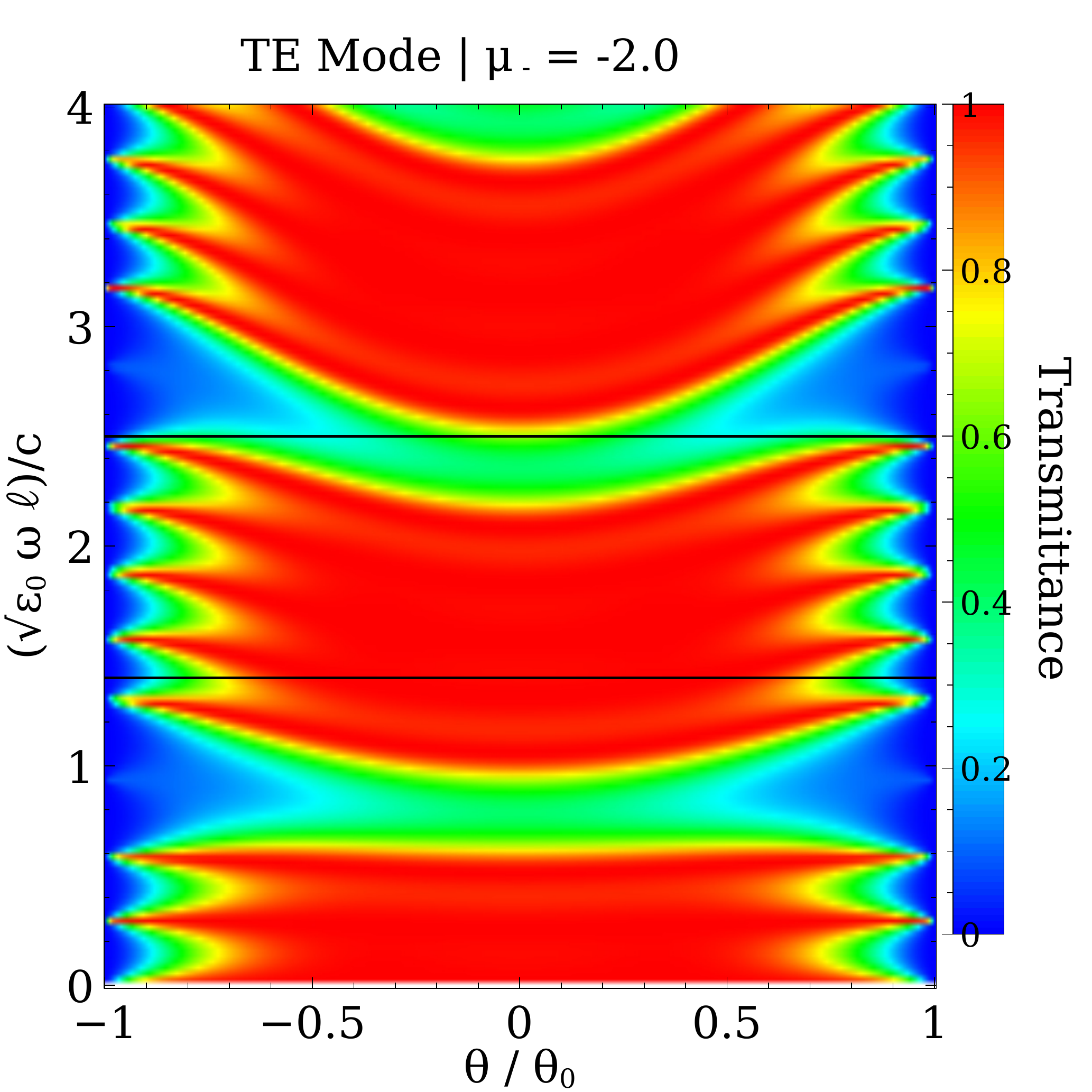}
\includegraphics[width=0.30\linewidth]{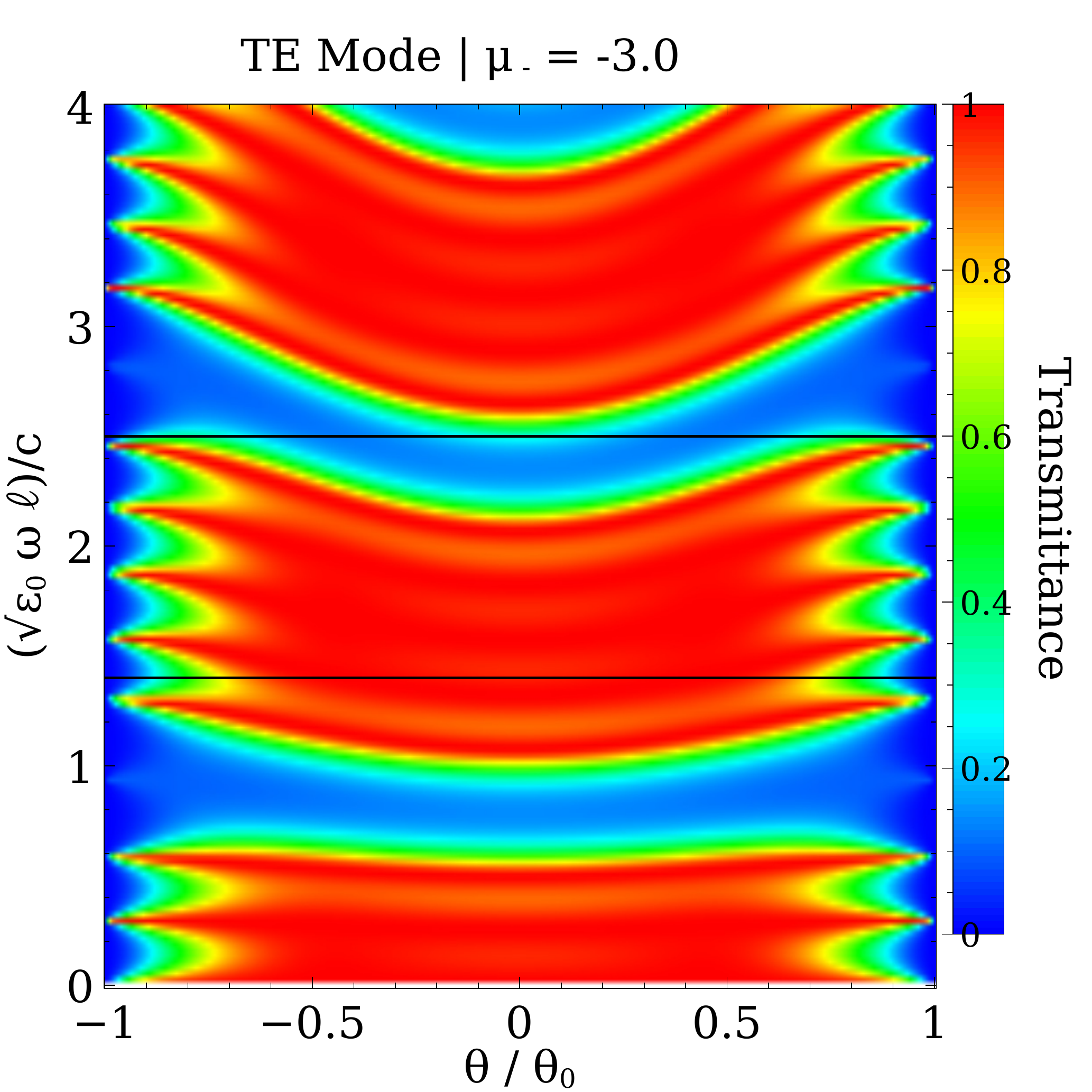}
\includegraphics[width=0.30\linewidth]{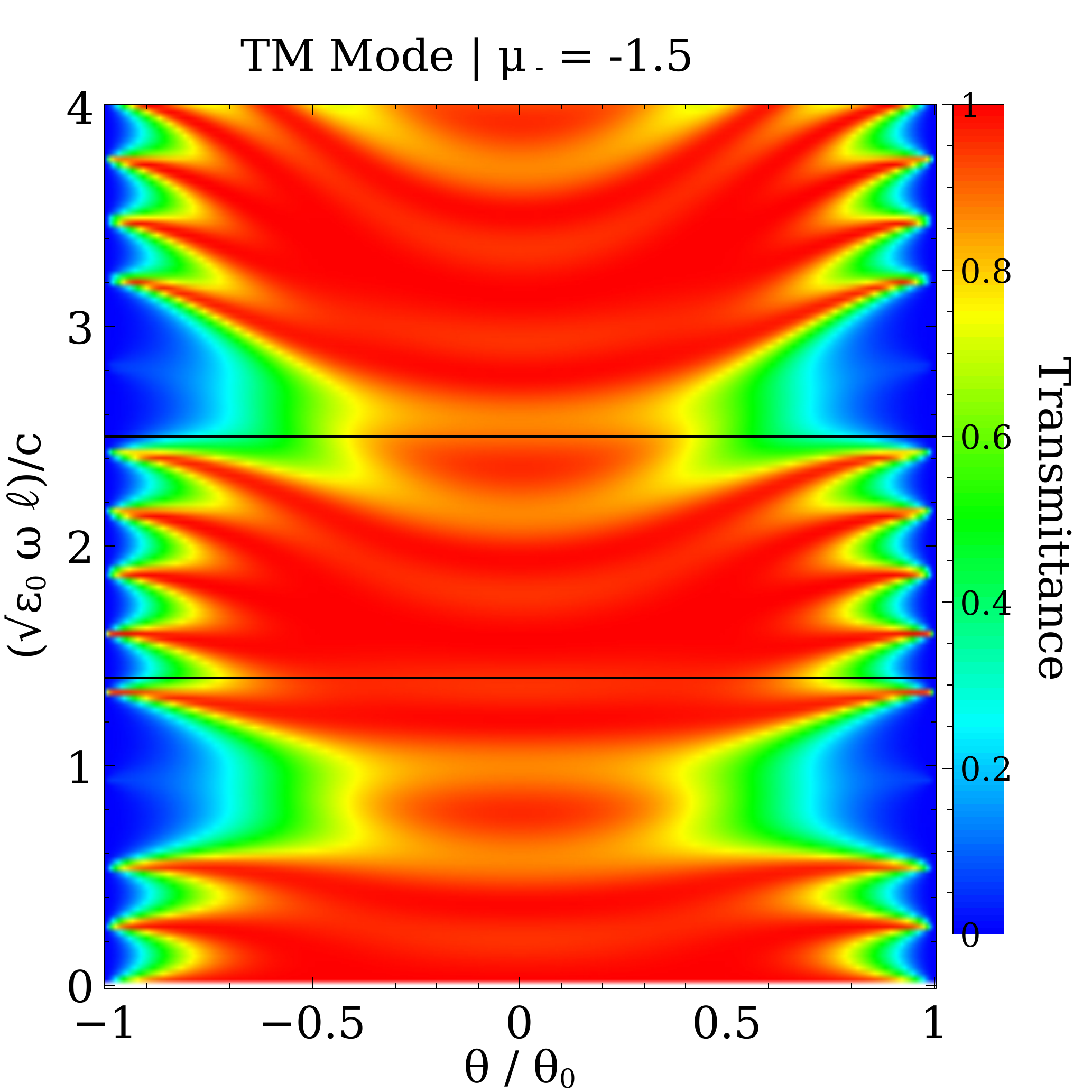}
\includegraphics[width=0.30\linewidth]{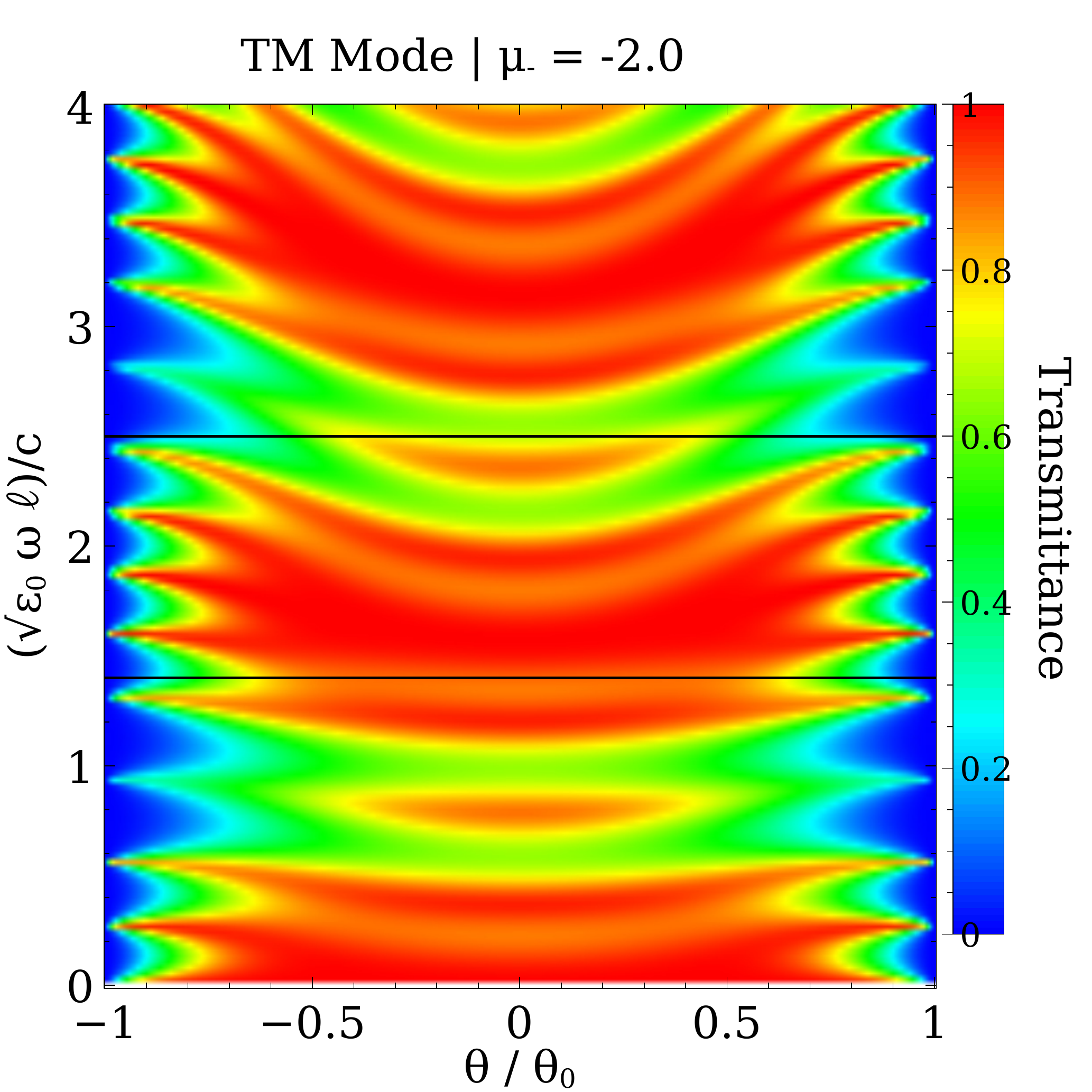}
\includegraphics[width=0.30\linewidth]{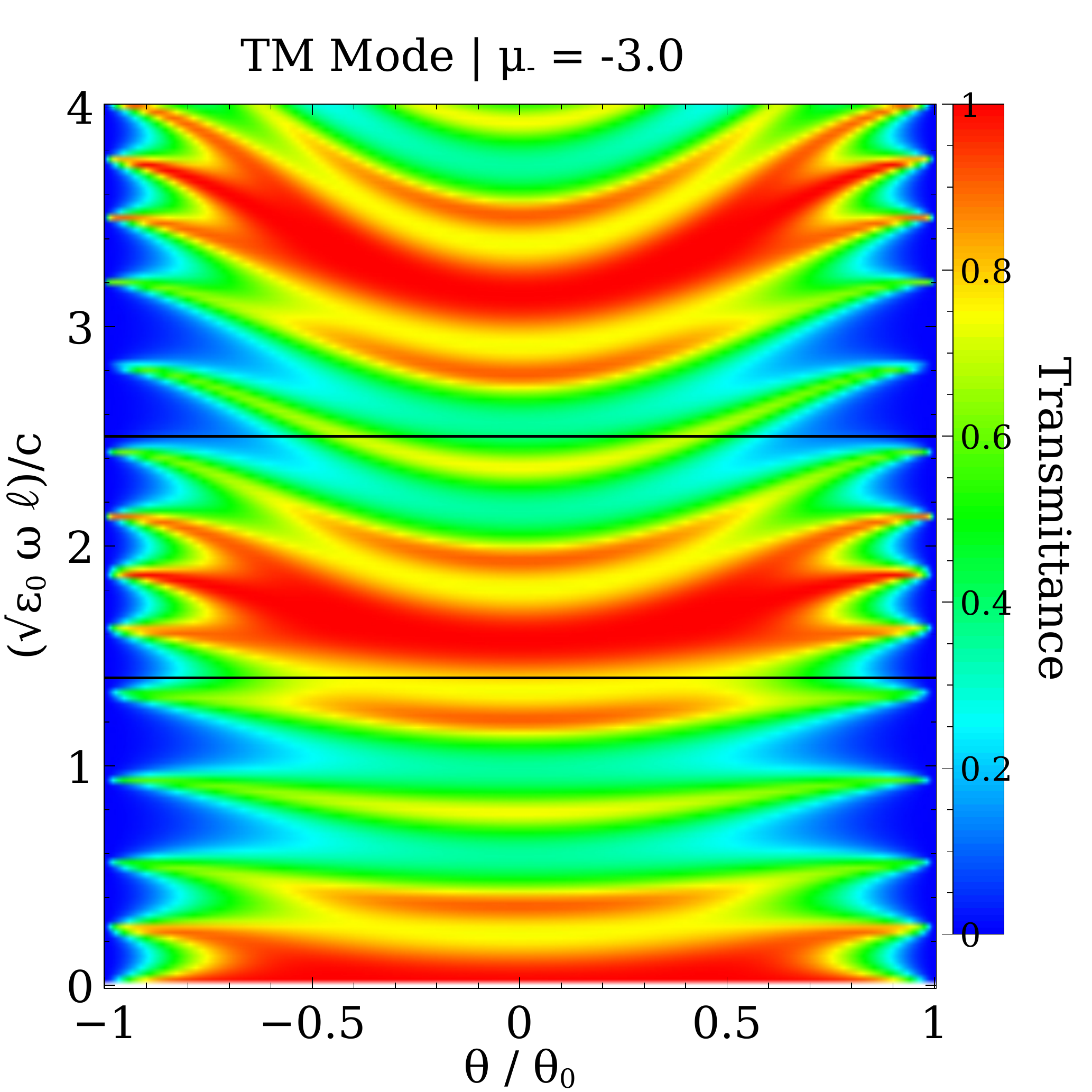}
\caption{Contour plots. The top row corresponds to the TE mode while the
bottom row corresponds to the TM mode. $\mu_+$ is fixed at 1.2.}
\label{mmv}
\end{figure*}

In the case of increasing the values of $|\mu_-|$, Fig.~(\ref{mmv}) shows
that the main effect is to diminish the transmittance between the three regions
of high transmittance. This effect is also visible on both the TE and TM modes. However, in this
case, we notice that there is no shifting effect, which can be seen easily from
the right panel of Fig.~(\ref{shift-m}).

\begin{figure*}[ht]\centering 
\includegraphics[width=0.30\linewidth]{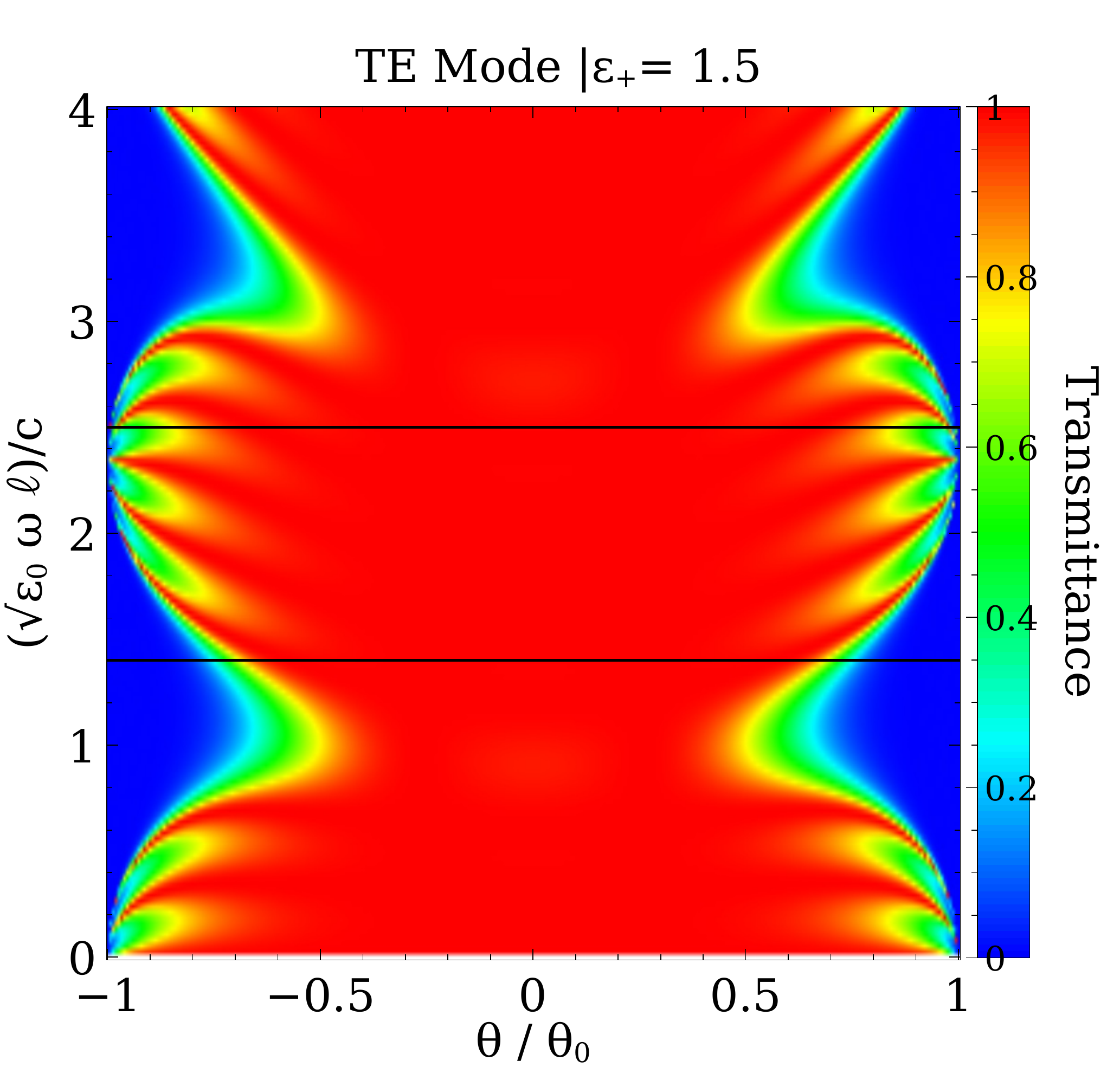}
\includegraphics[width=0.30\linewidth]{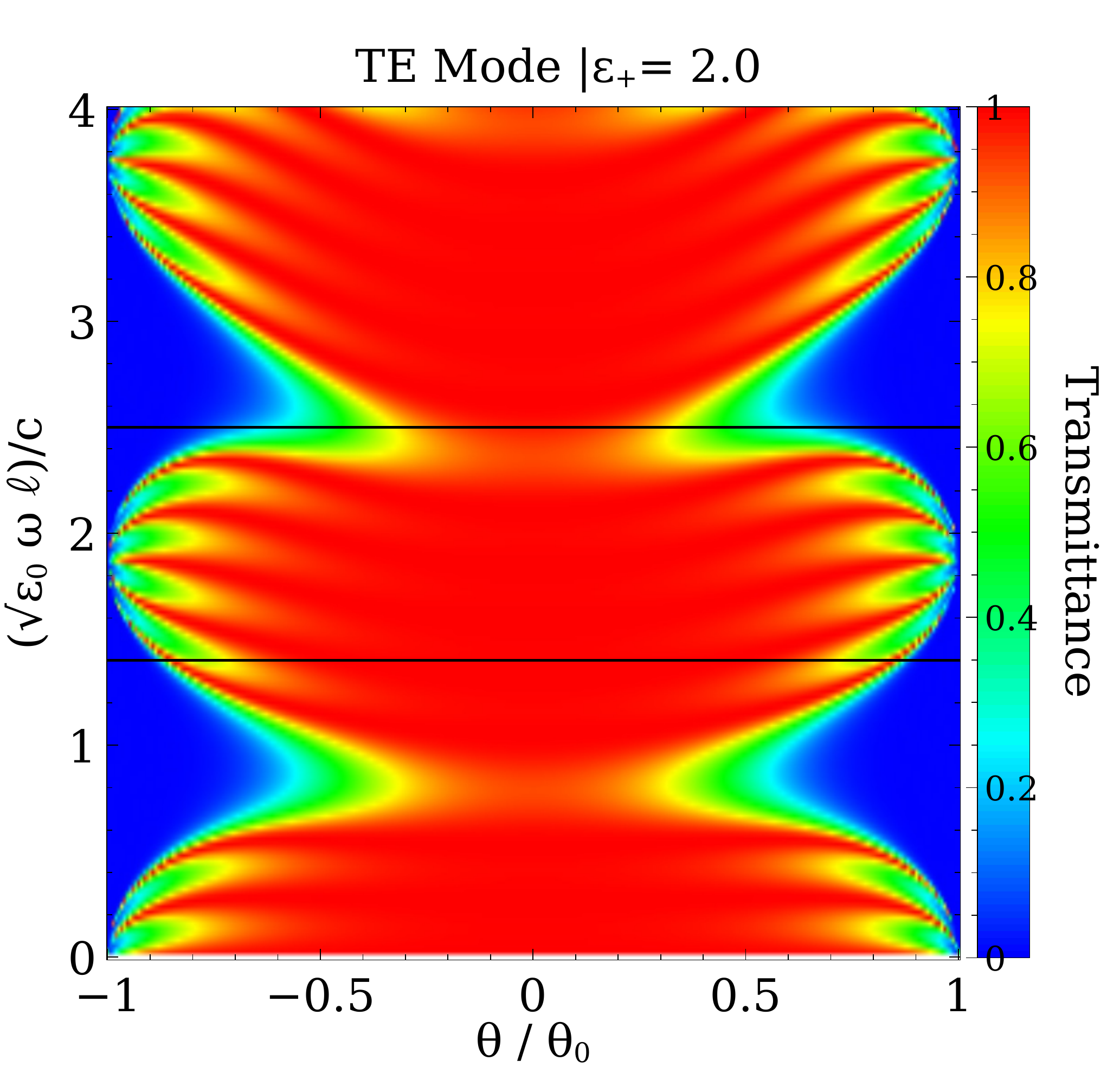}
\includegraphics[width=0.30\linewidth]{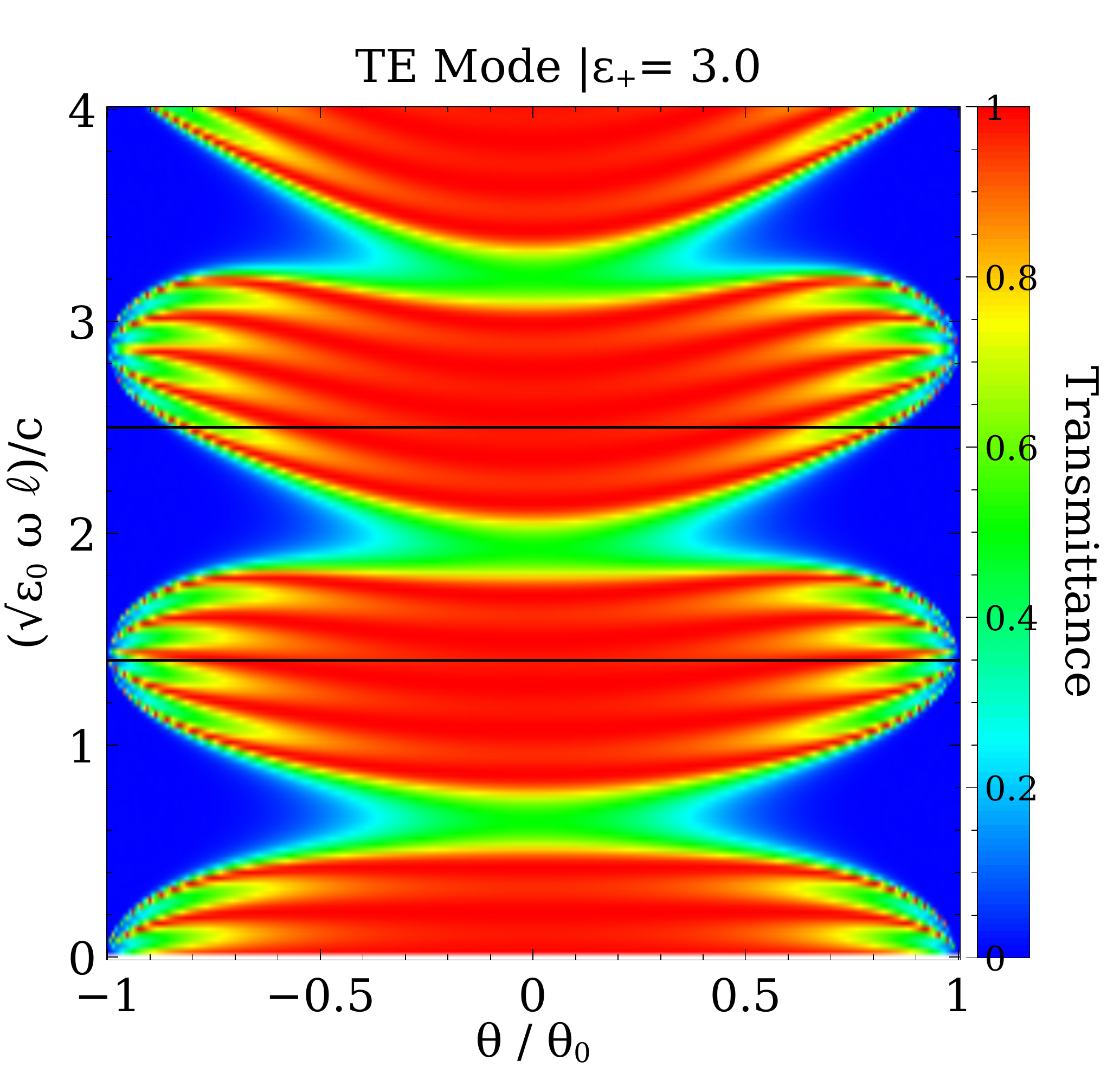}
\includegraphics[width=0.30\linewidth]{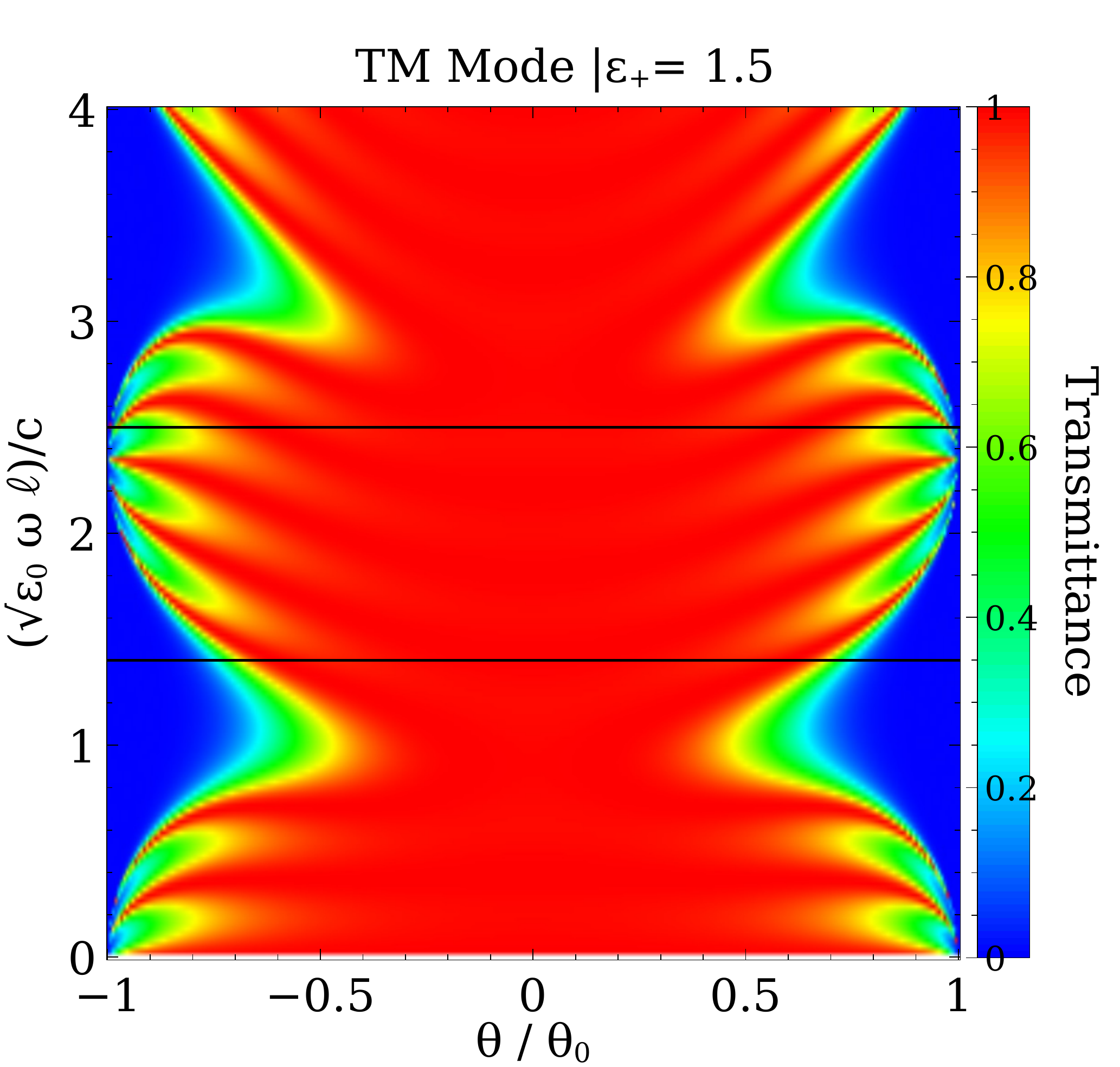}
\includegraphics[width=0.30\linewidth]{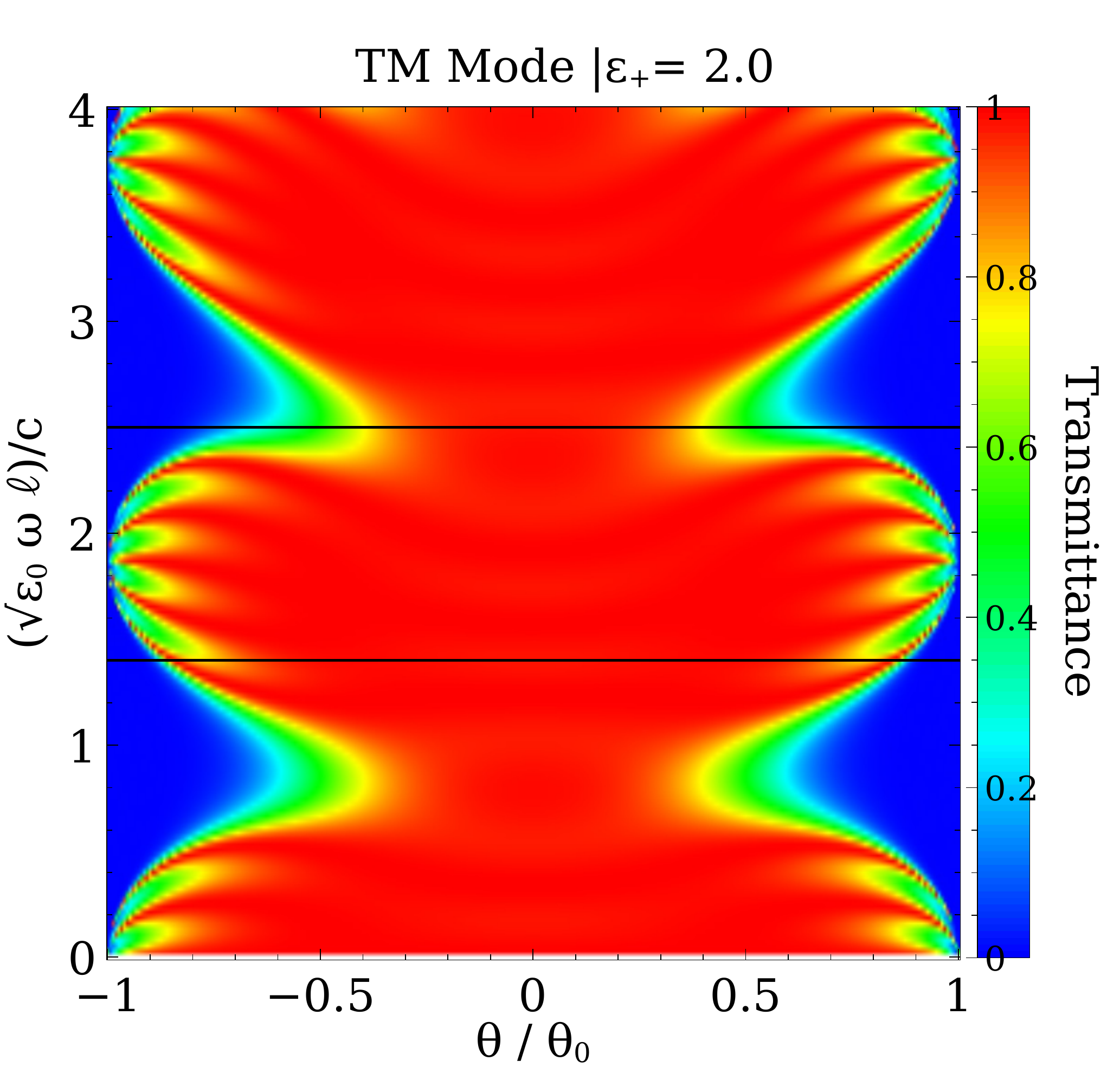}
\includegraphics[width=0.30\linewidth]{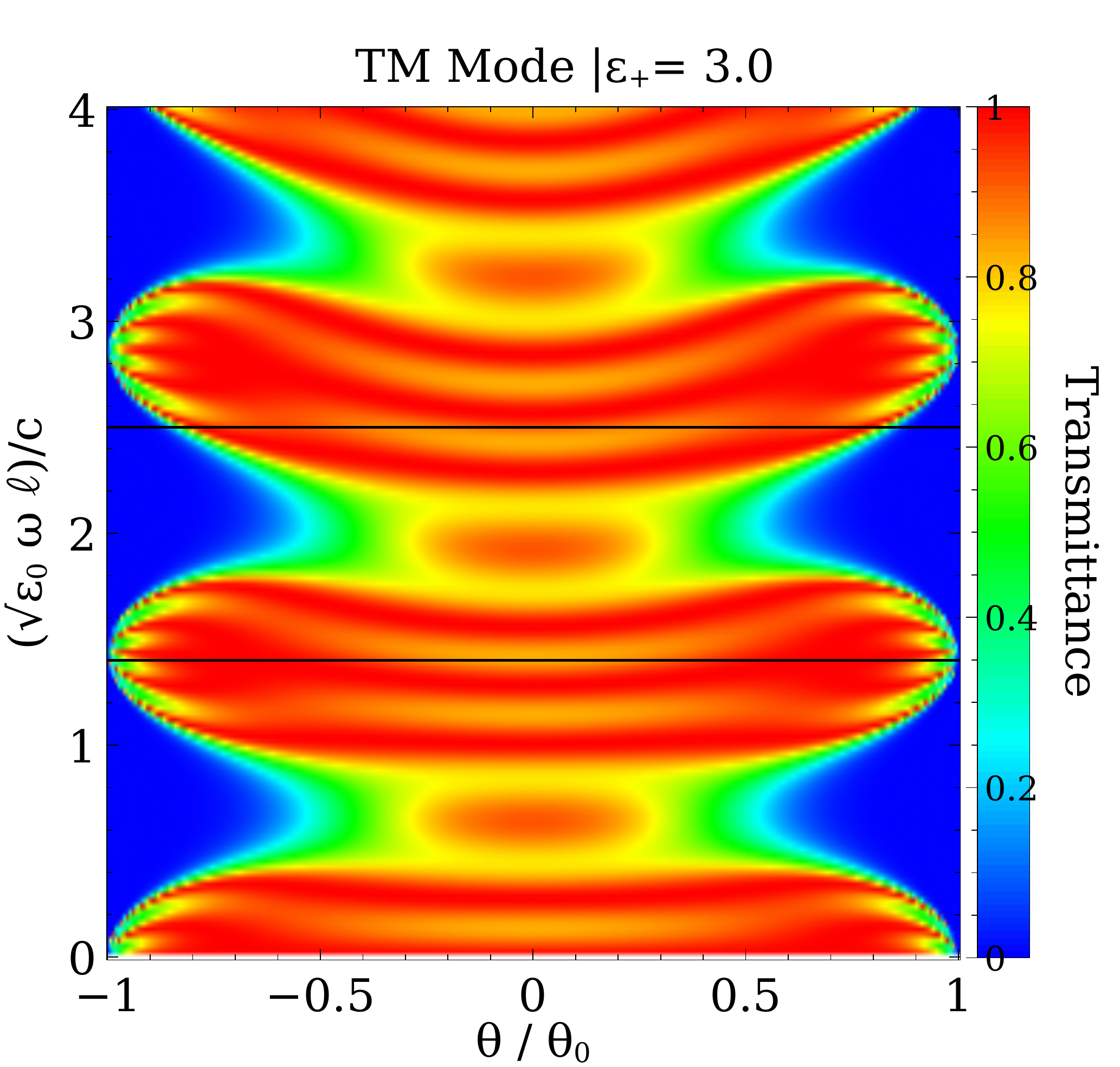}
\caption{Contour plots. The top row corresponds to the TE mode while the
bottom row corresponds to the TM mode. The positive permeability takes values
$\epsilon_+=1.5,2,3$ while $\epsilon_-$ is fixed at -1.0.}
\label{epv}
\end{figure*}

\begin{figure*}[ht]\centering 
\includegraphics[width=0.30\linewidth]{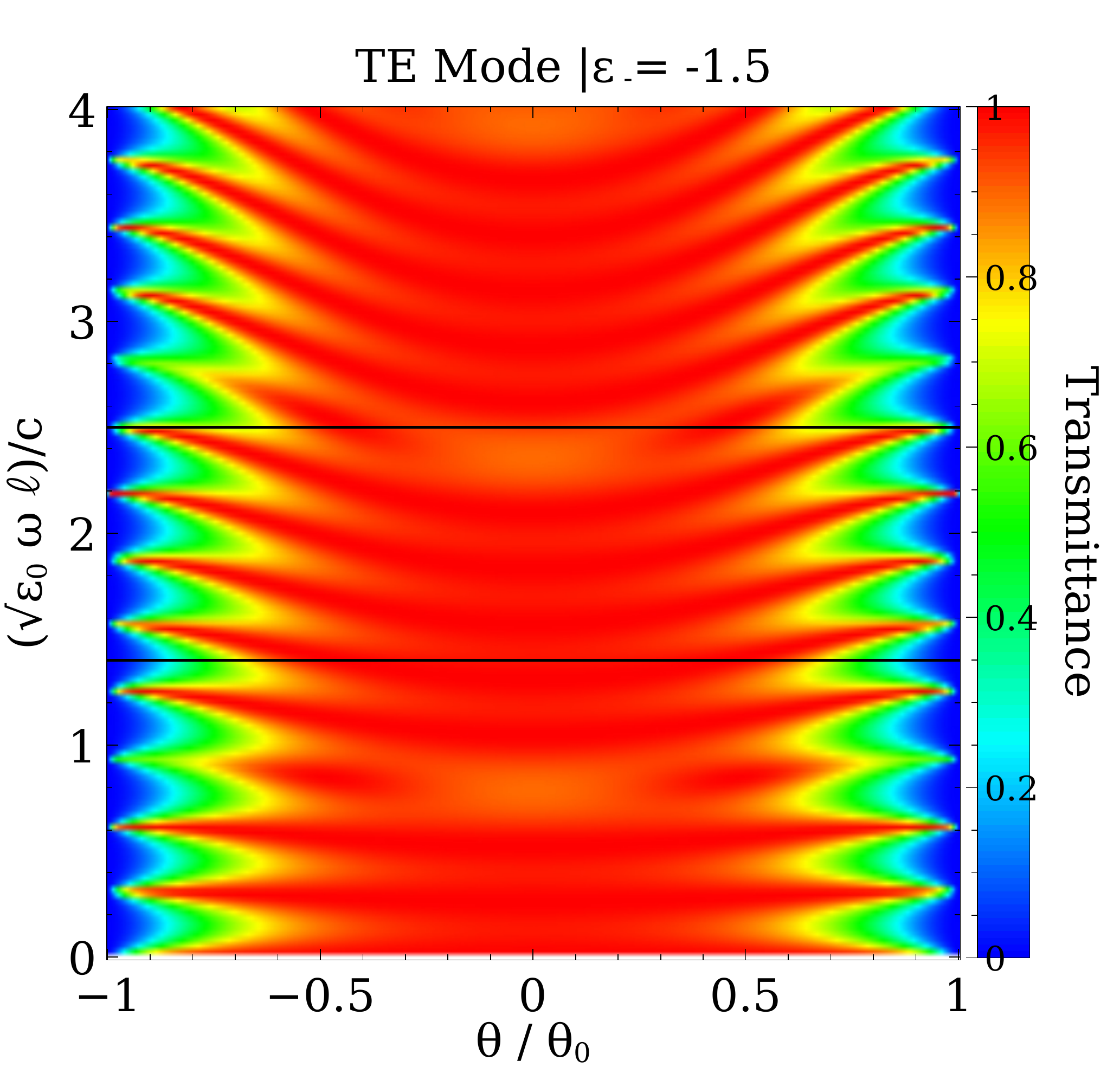}
\includegraphics[width=0.30\linewidth]{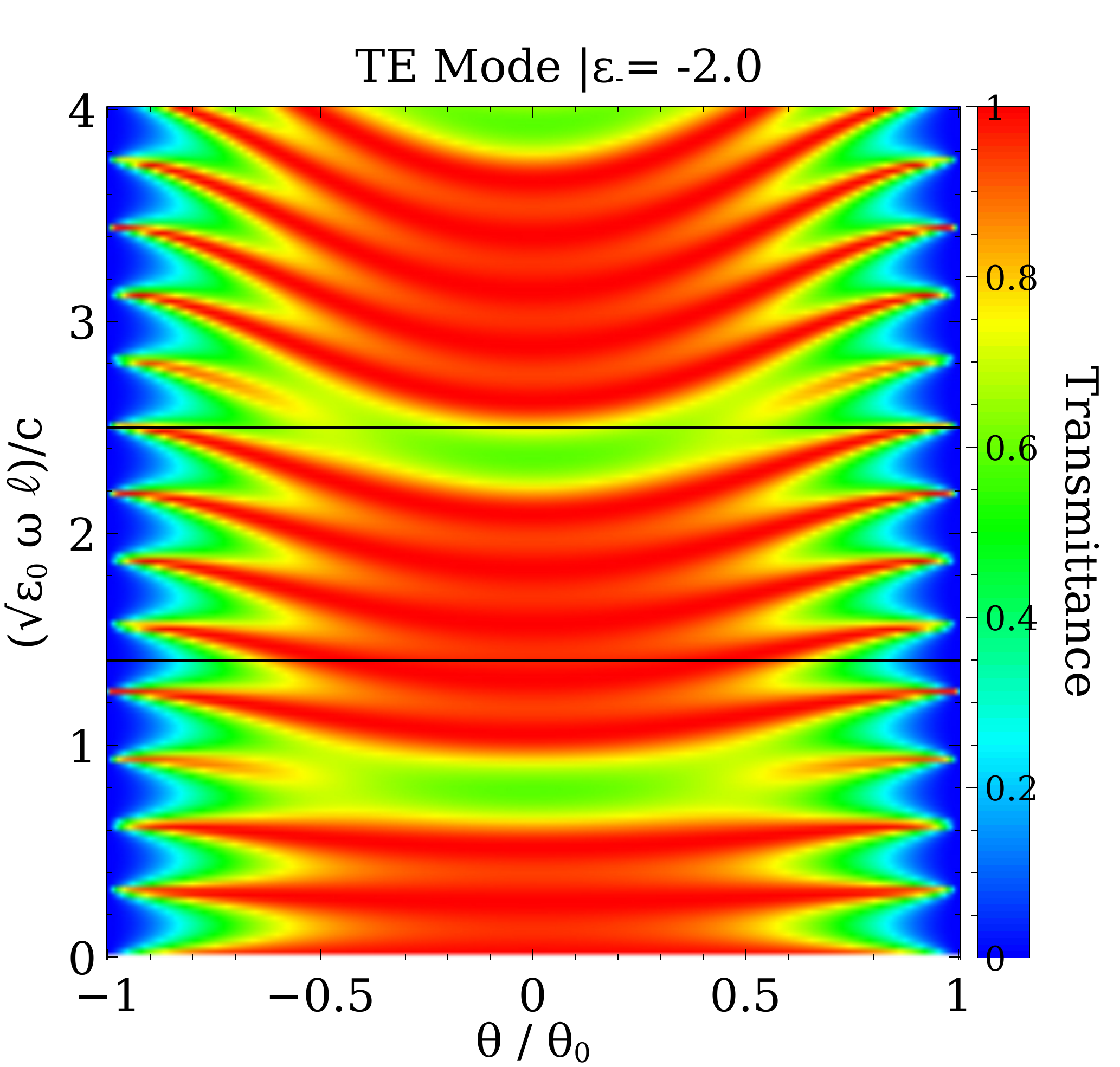}
\includegraphics[width=0.30\linewidth]{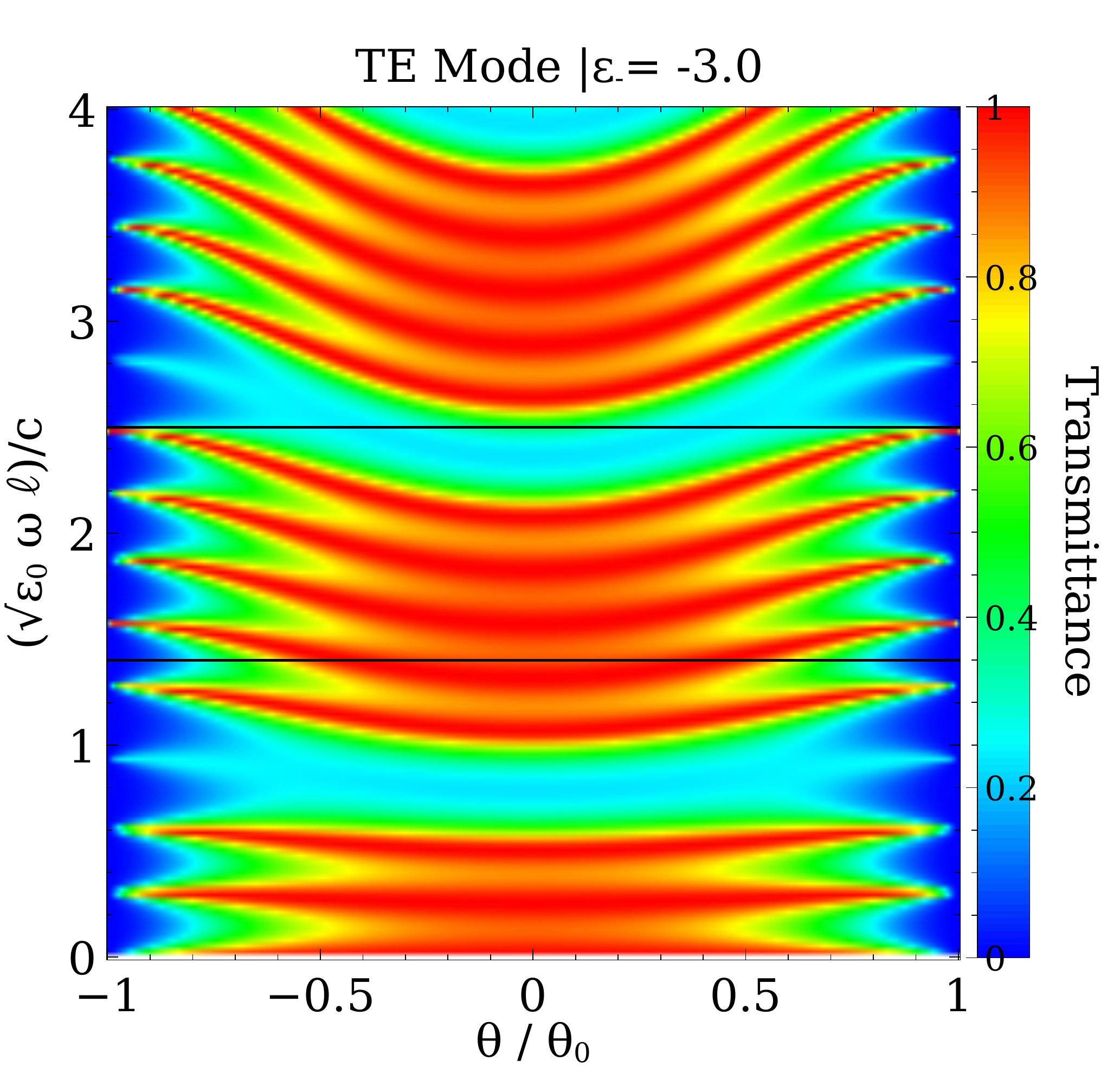}
\includegraphics[width=0.30\linewidth]{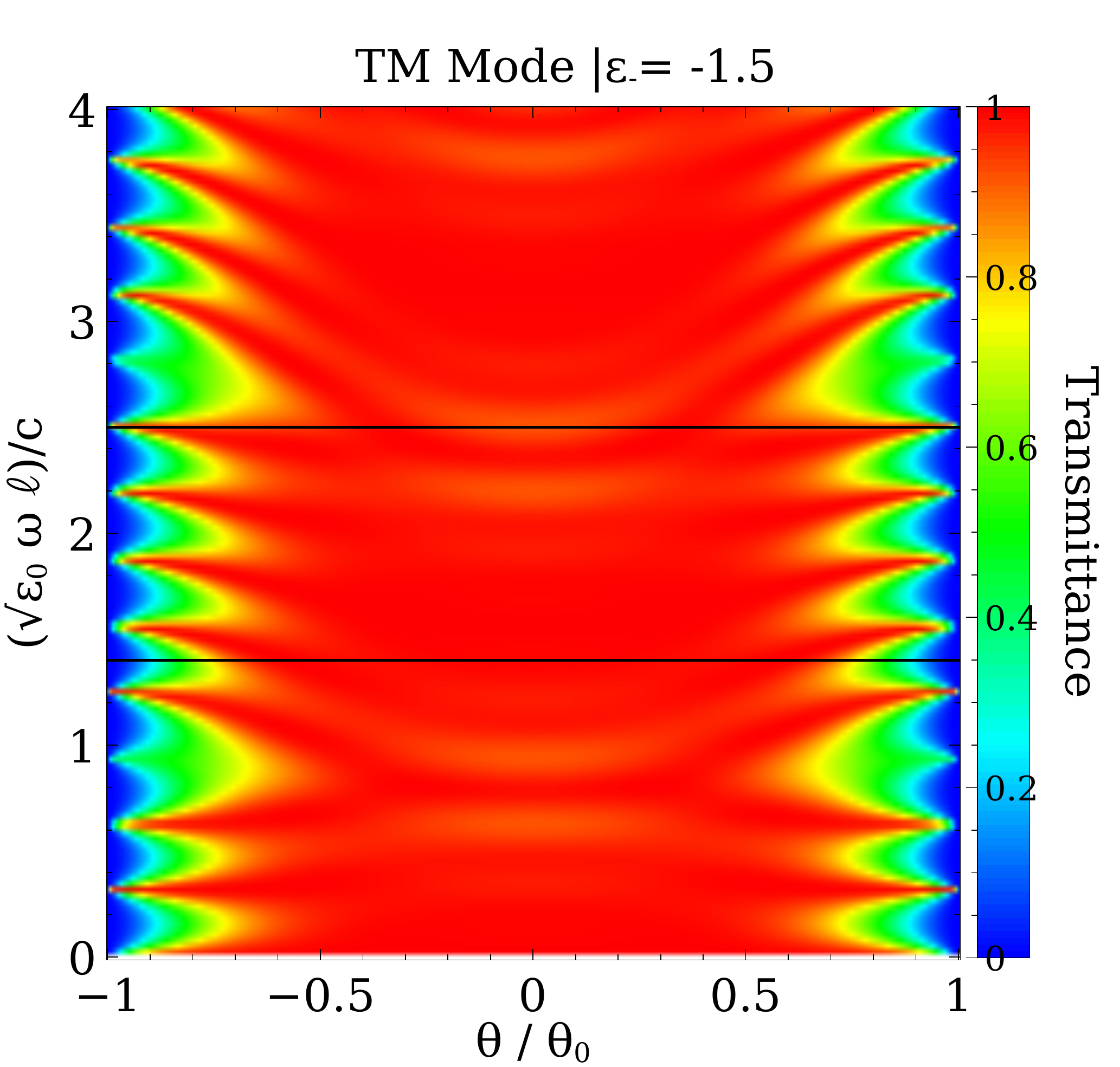}
\includegraphics[width=0.30\linewidth]{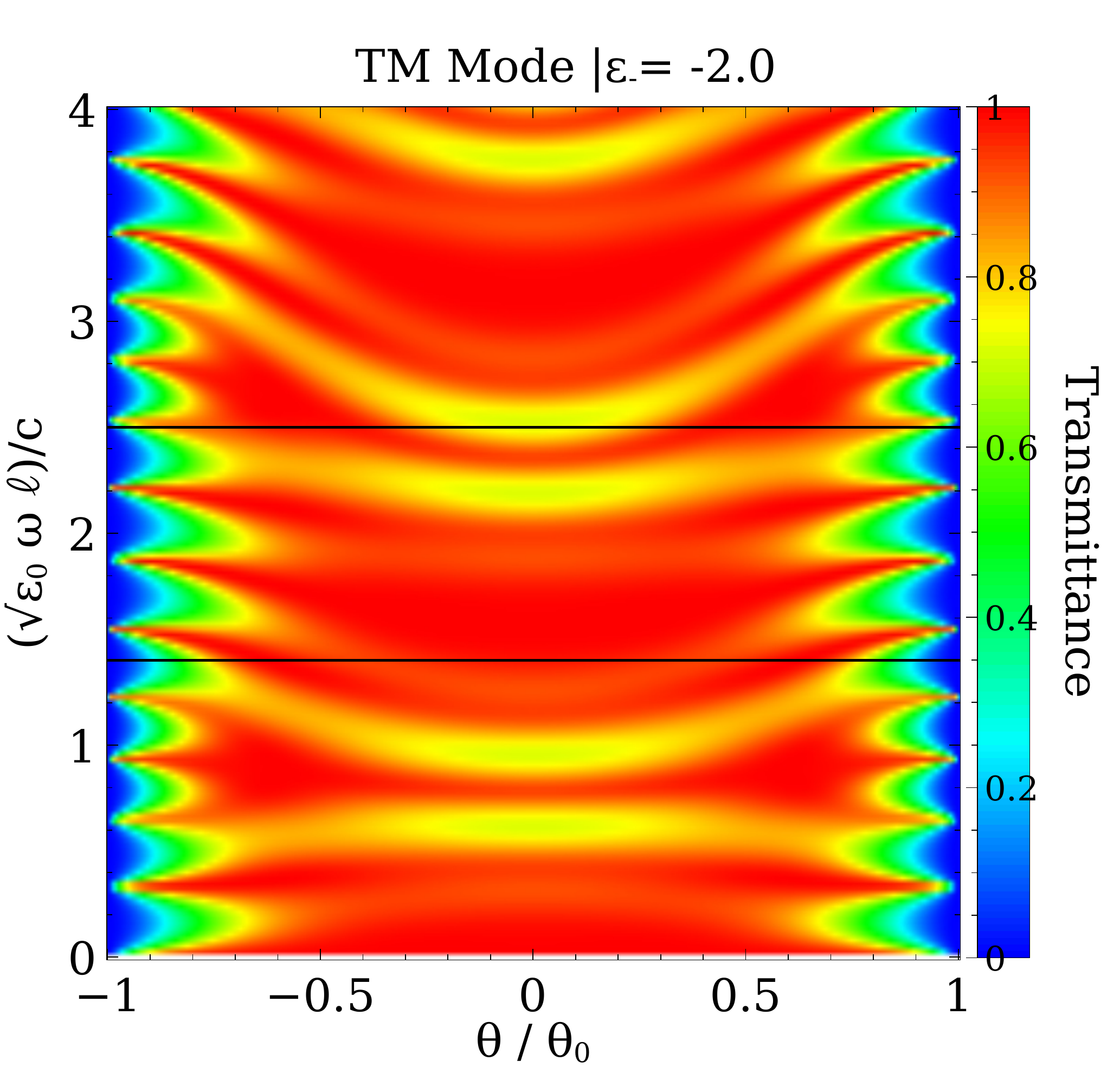}
\includegraphics[width=0.30\linewidth]{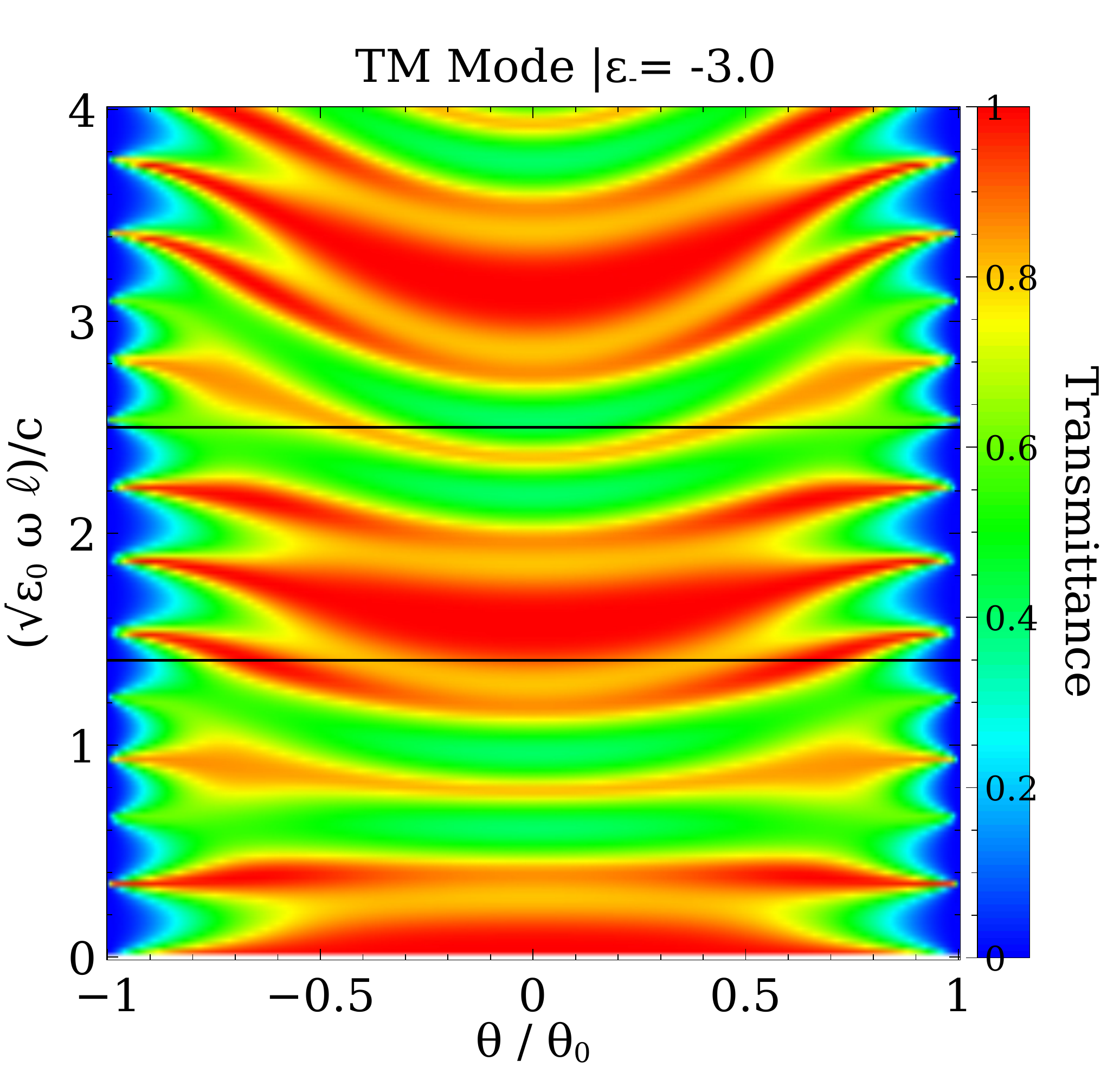}
\caption{Contour plots. The top row corresponds to the TE mode while the
bottom row corresponds to the TM mode. The positive permeability takes values
$\epsilon_-=1.5,2,3$ while $\epsilon_+$ is fixed at 2.0.}
\label{emv}
\end{figure*}

In addition, Figs.~(\ref{ctnb}) - (\ref{mmv}) show that the regions of high transmittance
do not vary with the angle of incidence, which may be related to the independence
of the polarization angle of the incident wave as also reported in \cite{Wang::2016}. Considering
that the reflectance and transmittance are complementary properties, our findings regarding
the transmittance spectrum seem to be in agreement with those reported in \cite{Shekhar::2014}
for an Ag-TiO$_2$ multilayer system, which is the standard multilayer candidate with high transmittance
in the visible region.

\section{Conclusion}

We have obtained the transmission properties of a multilayered structure of
alternating positive index media and negative index media using the TMM.
We have sought for profiles of high transmittance at any angle of incidence in the visible region of the spectrum.
In this respect, our results suggest a system with the properties displayed in Fig.~(\ref{ctnb}),
i.e., with $\epsilon_+=2.0,\epsilon_-=-1.0,\mu_+=2.0$, and $\mu_-=-1.2$, for the case of five blocks,
$n_b = 5$, which present an almost uniform region of high transmittance in the visible part
of the electromagnetic spectrum.

In addition, we have studied the manner in which the permeability and permittivity parameters
affect the transmission properties in an alternation of material and metamaterial. We have observed that positive
permeabilities may have the effect of reversing the appearance of the transmission bands that one may see
in multilayered structures. On the contrary, positive permittivities have the effect of making the transmittance bands
more pronounced. Somewhat surprisingly, the negative values of the electromagnetic material parameters do not have relevant effects
in these settings.

The results obtained in this paper are valid far from the absorption bands of the multilayer structure to avoid numerical instability problems which are known to occur when the absorption is included \cite{hsueh}. However, the results we report for the visible region may still remain valid for some metamaterials, such as the hyperbolic ones, for which the absorption bands lie in the ultraviolet region \cite{tum}.

\section*{Acknowledgments} 

\addcontentsline{toc}{section}{Acknowledgments} 


The first author acknowledges the financial support of CONACyT through a doctoral fellowship at IPICyT.


\begin{thebibliography}{10}
%
%
%
%
%
%
%

\bibitem{Banerjee::2011} P.P. Banerjee, H. Li, R. Aylo, and G. Nehmetallah, Transfer matrix approach to propagation of angular plane wave spectra through metamaterial multilayer structures, {\em Proc. SPIE} {\bf 8093} (2011) 3387-3389.

\bibitem{Solymar::2009} L. Solymar and E. Shamonina, {\em ``Waves in Metamaterials''} (Oxford University Press, 2009).

\bibitem{Markos::2008} P. Markos and C.M. Soukoulis, {\em ``Wave propagation from electrons to photonic crystals and left-handed materials''}
(Princeton University Press, 2008).

\bibitem{python} Python Software Foundation. Python Language Reference, version 2.7. Available at http://www.python.org.


\bibitem{Wang::2016} B.-X. Wang, Single-patterned metamaterial structure enabling multi-band perfect absorption,
{\em Plasmonics} {\bf 12} (2017) 95-102. 

\bibitem{Shekhar::2014} P. Shekhar, J. Atkinson, and Z. Jacob, Hyperbolic metamaterials: fundamentals and applications, {\em Nano Convergence} {\bf 1} (2014) 14.

\bibitem{hsueh} W.-J. Hsueh and J.-C. Lin, Stable and accurate method for modal analysis of multilayer waveguides using a graph approach,
{\em J. Opt. Soc. Am.} {\bf 24} (2007) 825-830.


\bibitem{tum} T. Tumkur, Y. Barnakov, S.T. Kee, M.A. Noginov, and V. Liberman,
Permittivity evaluation of multilayered hyperbolic metamaterials. Ellipsometry vs. reflectometry, {\em J. Appl. Phys.} {\bf 117} (2015) 103104.

\end{thebibliography}
\end{document}